\documentclass[10pt,twocolumn]{IEEEtran}
\pdfoutput=1
\usepackage{graphicx}
\usepackage{amsmath}
\usepackage{amssymb}
\usepackage[noadjust]{cite}
\usepackage{float}
\usepackage{color}
\usepackage{url}
\usepackage{dblfloatfix} 
\usepackage{xfrac}
\usepackage{amsfonts}
\usepackage{enumitem}
\usepackage{float}
\usepackage{color}
\usepackage{cite}
\usepackage{graphicx}
\usepackage{epstopdf}
\usepackage{bm}
\usepackage{subcaption}
\usepackage{relsize}
\usepackage{mathtools}
\usepackage{array}
\usepackage{array}
\usepackage{multirow}
\usepackage{multicol}
\usepackage{mathtools}
\usepackage{blindtext}
\usepackage{wrapfig}
\usepackage{cuted}
\usepackage{lipsum}
\usepackage{xcolor}
\usepackage{amssymb}
\usepackage{pifont}
\usepackage{makecell}

\newtheorem{Remark}{Remark}

\allowdisplaybreaks

\graphicspath{{Figures/}}
\begin{document}
\title{Analysis of UAV Radar and Communication Network Coexistence with Different Multiple Access Protocols}
\author{Sung Joon Maeng, Jaehyun Park,~\IEEEmembership{Member, IEEE}, and \.{I}smail G\"{u}ven\c{c},~\IEEEmembership{Fellow, IEEE} \\
\thanks{This work has been supported in part by the NSF award CNS-1910153. This work has been also supported in part by National Research Foundation of Korea under the framework of international cooperation program (2022K2A9A2A06035926).}\thanks{S. J. Maeng, \.{I}. G\"{u}ven\c{c} are with the Department of Electrical and Computer Engineering, North Carolina State University, Raleigh, NC 27606 USA (e-mail: smaeng@ncsu.edu; iguvenc@ncsu.edu).} \thanks{J. Park with the Department of Electronic Engineering, Pukyong National
University, Busan 608-737, South Korea (e-mail: jaehyun@pknu.ac.kr).}
}

\maketitle
\begin{abstract}
Unmanned aerial vehicles (UAVs) are expected to be used extensively in the future for various applications, either as user equipment (UEs) connected to a cellular wireless network, or as an infrastructure extension of an existing wireless network to serve other UEs. Next generation wireless networks will consider the use of UAVs for joint communication and radar and/or as dedicated radars for various sensing applications. Increasing number of UAVs will naturally result in larger number of communication and/or radar links that may cause interference to nearby networks, exacerbated further by the higher likelihood of line-of-sight signal propagation from UAVs even to distant receivers. With all these, it is critical to study  network coexistence of UAV-mounted base stations (BSs) and radar transceivers. In this paper, using stochastic geometry, we derive closed-form expressions to characterize the performance of coexisting UAV radar and communication networks for spectrum overlay multiple access (SOMA) and time-division multiple access (TDMA). We evaluate successful ranging probability (SRP) and the transmission capacity (TC) and compare the performance of TDMA and SOMA. Our results show that SOMA can outperform TDMA on both SRP and TC when the node density of active UAV-radars is larger than the node density of UAV-comms.
\end{abstract}


\section{Introduction}\label{sec:intro}
Recently, various different applications of cellular-connected unmanned aerial vehicles (UAVs) have been getting significant attention due to their cost-efficient deployment and controllable mobility. UAVs are utilized in many fields such as environmental monitoring and surveillance~\cite{7463007}, public safety~\cite{7122576}, video broadcasting~\cite{8214963}, and delivery~\cite{8565971}. Moreover, UAV-mounted base stations (UAV-BS) and user equipment (UAV-UE), as well as UAV-mounted radars (UAV-radar) are commonly considered in sensing and communication applications, since the UAVs are available to quickly change the position to serve users at the outage area, and/or surveil/track the location of detected moving targets. Joint design of the radar and communication systems is considered as one of the key research areas for wireless networks beyond 5G systems which can benefit significantly from the use of autonomous UAVs.

In the meanwhile, as the demand for using wider bandwidths has been increasing to support higher throughput and massive connectivity, the band of operation for broadband wireless networks has been moving to higher frequencies such as millimeter-wave (mmWave) and sub/THz bands that are also commonly used by radar systems. Some traditional radar bands, including certain bands below 6~GHz, are also being opened for shared use with communication networks due to the increasing congestion in the dedicated spectrum for cellular networks. All these developments call for rigorously studying the coexistence scenarios for radar and communication networks and coming up with strategies for effective spectrum sharing~\cite{8828016}.

Stochastic geometry-based techniques are commonly used in the literature for obtaining closed-form expressions on the performance of wireless networks where transmit sources are randomly deployed in the spatial domain~\cite{7733098}. 
For example, the analysis of accumulated interference from multiple nodes following a homogeneous Poisson point process (HPPP) is useful to evaluate the capacity of wireless networks~\cite{7733098}. In this paper, we specifically investigate UAV radar sensing and communication network coexistence scenarios. In particular, we consider scenarios where radar transmission and data transmission are coordinated by two different multiple-access protocols: spectrum overlay multiple access (SOMA) and time-division multiple access (TDMA). In SOMA, radar sensing and data communication share the same spectrum so that the spectrum is overlapped. On the other hand, in TDMA, radar detection and communication are separated by time. We utilize stochastic geometry-based analysis where UAVs are randomly located in 3D space following a two-dimensional homogeneous Poisson point process (HPPP). We individually analyze the radar detection performance and the data communication performance using the successful ranging probability (SRP) and the transmission capacity (TC), respectively.

\begin{table*}[t!]
\renewcommand{\arraystretch}{1.1}
\caption{Literature review for stochastic geometry based wireless network performance analysis.}
\label{table:settings}
\centering
\begin{tabular}{clccl}
\hline
Ref. &  Analysis objective & Application & Radar & Networks \\
\hline\hline
\cite{4712724} &   TC with different spatial diversity techniques  & Terrestrial & \ding{55} & Ad hoc \\
\cite{8106812} &   Information and energy outage probability and area harvested energy & Terrestrial & \ding{55} & SWIPT in ad hoc networks\\
\cite{9043513} &   Channel outage and packet loss probability & Terrestrial & \ding{55} & DL URLLC communications\\
\cite{7819520,9119440} &   SRP & Terrestrial & \checkmark & Radar with road scenario \\
\cite{8464057} &   SRP & Terrestrial & \checkmark & Radar networks  \\
\cite{8628991,7913628} &   Coverage probability with different antenna patterns and directivity & Terrestrial & \ding{55} & DL cellular \\
\cite{7464352} &   Connection and secrecy probability & Terrestrial & \ding{55} & DL secure communication cellular \\
\cite{9115898} &   Coverage probability with underlay and overlay protocols & UAV & \ding{55} & UAV-to-UAV and UL terrestrial\\
\cite{8876665} &   Coverage probability and spectral efficiency  & UAV & \ding{55} & Two-tiers cellular\\
\cite{7967745,9035640} &   Coverage probability and spectral efficiency & UAV & \ding{55} & DL communication\\
\cite{8876702} &   Successful transmission probability, energy and SINR coverage & UAV & \ding{55} & DL SWIPT and UL communication\\
\cite{8856258} &   Coverage probability & UAV & \ding{55} & UAV-aided DL and UL communication\\
\cite{9419751} &   LoS probability & UAV & \ding{55} & BS-to-UAV link\\
\cite{9031752} &   Coverage probability, motion energy, and flight time & UAV & \ding{55} & UAV path planning\\
\cite{8918344} &   Connection, secrecy, and energy-information outage probability & UAV & \ding{55} & Secure communication in SWIPT\\
This work &   SRP and TC & UAV & \checkmark & Radar and communication coexistence\\
\hline
\hline
\end{tabular}
\end{table*}

Contributions of this paper can be summarized as follows:
\begin{itemize}
  \item[--] We derive closed-form expressions for SRP and TC on the UAV radar and communication coexistence scenario where UAVs are placed following HPPP with a guard zone.
  \item[--] We analyze the performance of  SRP and TC in SOMA and TDMA, respectively. We also investigate behaviors of  SRP and TC depending on the node density, radius of the guard zone, power splitting factor in SOMA, and time division factor in TDMA.
  \item[--] We analytically compare TDMA and SOMA on  SRP and TC and show that TDMA outperforms SOMA on SRP while SOMA is better than TDMA on TC in the general condition. Furthermore, we analyze the condition that SOMA can be superior to TDMA on both  SRP and TC metrics.
\end{itemize}
The rest of this paper is organized as follows. Section~\ref{sec:literature_review} presents the literature review. In Section~\ref{Sec:Sys_Model}, we describe the UAV radar and communication network coexistence  design. In Section~\ref{Sec:HPPP}, we provide the signal propagation model when UAVs are distributed by HPPP. In Section~\ref{Sec:Perform_ana}, we derive the closed-form expressions of the SRP and the TC. In Section~\ref{Sec:SRP_ana}, we analyze SRP depending on the system parameters and the multiple access protocols. In Section~\ref{Sec:TC_ana}, we analyze the TC depending on the system parameters and the multiple access protocols. In Section~\ref{Sec:SOMA_TDMA}, we compare the SRP and the TC performance of SOMA and TDMA. In Section~\ref{Sec:Results}, we show the simulation results to verify the analysis in previous sections, and Section~\ref{Sec:Conclusion} provides concluding remarks. 

\section{Literature Review}\label{sec:literature_review}
The operation of UAVs on BSs and radar detectors have been investigated in the literature. A flying UAV-BS can maximize the capacity or minimize the outage of networks by optimizing UAV trajectory~\cite{9037325,9367288}. In \cite{8392472,9154268}, the trajectory and precoder of UAV-BS are optimized to maximize physical layer secrecy. In \cite{8641294}, a UAV-radar is used in measuring the depth of the snow on the sea. Human detection and classification by a UAV-radar have been studied in \cite{7094509}. Target detection using radar imaging from UAV-radar has been investigated in \cite{7991432}. In \cite{8580940}, the feasibility of a surveillance system using a UAV-radar has been explored. 

The study of coexistence networks has been explored in the literature. In \cite{7898445}, a beamforming approach has been studied to facilitate the coexistence between downlink (DL) multi-user-multiple-input-multiple-output (MU-MIMO) communication and MIMO radar system. In \cite{8335405}, the joint design of the radar and communication system for the coexistence of MIMO radar and MIMO communication has been studied. Moreover, UAV communication and radar sensing network coexistence that utilizes the spectrum for both purposes has been investigated for an efficient and flexible system design~\cite{9456851}. In \cite{9282206}, joint UAV communication and cooperative sensing network based on beam sharing scheme has been explored.

Stochastic geometry based network analysis has been thoroughly investigated in the literature. TC is analyzed in ad hoc networks with different spatial diversity techniques where transmitting nodes are distributed by an HPPP~\cite{4712724}, and this work is extended to the wireless information and power transfer (SWIPT)-based ad hoc networks in~\cite{8106812}. In \cite{9043513}, packet loss probability depending on the packet size, packet duration, and SINR are derived in downlink ultra-reliable and low-latency communications (URLLC) scenarios where distributed antenna ports are randomly placed following an HPPP. In~\cite{7819520,9119440,8464057}, the effect of radar interference on the radar detection performance is analyzed and SRP is evaluated using stochastic geometry. More specifically, the geometric layout of vehicles on a road where the locations of vehicles on a certain lane are decided by unidimensional HPPP model is investigated in~\cite{7819520,9119440}.
In~\cite{9119440}, radar cross-section (RCS) characteristics are modeled and analyzed using HPPPs for automotive radar network scenarios.  In~\cite{8628991,7913628}, the effects of different directional antenna patterns, node densities, and antenna array sizes on coverage probability are studied for mmWave networks. The locations of BSs and the eavesdroppers are randomly  distributed by independent HPPPs in~\cite{7464352}, and  closed-form expression of secrecy probability for secure communications is explored.

Closed-form analysis of network performance using stochastic geometry techniques have also been studied in UAV networks in~\cite{9115898,8876665,7967745,9035640,8876702,8856258,9419751,9031752,8918344}. In~\cite{9115898}, coexisting UAV-to-UAV links and uplink (UL) ground-BS to ground-user links are considered. Then, coverage of two different scenarios are studied, where the spectrum for each link is either reused, or it is allocated in a dedicated manner. The literature review with representative works related to stochastic geometry-based wireless network performance  analysis is summarized in Table~I. To the best of our knowledge, the study of radar networks based on stochastic geometry has been limited, and UAV communication and radar network coexistence scenario has not been investigated yet.

\begin{table}[!t]
\renewcommand{\arraystretch}{1.1}
\caption{Key symbols and notations used in this paper.}
\label{table:settings}
\centering
\begin{tabular}{ll}
\hline
Symbol & Definition \\
\hline\hline
${\lambda_{\rm r}}^{\prime}$ & Node density of UAV-radars\\
${\lambda_{\rm d}}^{\prime}$ & Node density of UAV-comms\\
${\bar\lambda}_{\rm r}$ & Active node density of UAV-radars\\
${\lambda_{\rm r}}$ & Effective node density of active UAV-radars in HPPP\\
${\lambda_{\rm d}}$ & Effective node density of UAV-comms in HPPP\\
$\mathsf{h}_{\rm UAV}$ & UAVs height\\
$r_0$  & Radius of guard zone\\
$\phi$  & Power splitting factor\\
$\tau$  & Time division factor\\
$\delta$  & Duty circle\\
$\mathsf{P}_{\rm Tx}$ & Transmit power\\
$\mathsf{G}_{\rm t}$ & Tx antenna gain\\
$\alpha$ & Path-loss exponent\\
$\alpha_{\rm I}$ & Path-loss exponent from the interference\\
$\mathsf{G}_{\rm r}$ & Receiver antenna gain\\
$\mathsf{G}_{\rm rI}$ & Receiver antenna gain from the interference\\
$\bar{\sigma}$ & Average RCS\\
$\sigma$ & RCS\\
$S_{\rm e}$ & Effective aperture of radar receiver\\
$\mathsf{G}_{\rm p}$ & Processing gain of radar receiver \\
$f_{\rm c}$ & Carrier frequency\\
$R_0$ & Target distance\\
$r_i$ & Distance from the interferer\\
$c$ & Speed of light\\
$h_0$ & Small-scale fading\\
$h_i$ & Small-scale fading from the interference\\
$\beta_{\rm th}$ & Target SINR threshold for outage probability\\
$\gamma{\rm th}$ & Target SINR threshold for successful range probability\\
$\beta_{0}$ & SINR of the received data signal\\
$\gamma{0}$ & SINR of the received radar signal\\
$N_0$ & Noise power\\
\hline
\hline
\end{tabular}
\end{table}

\section{System Model}\label{Sec:Sys_Model}
\begin{figure}[t]
    \centering
    \subfloat[Radar detection scenario.]{\includegraphics[width=0.45\textwidth]{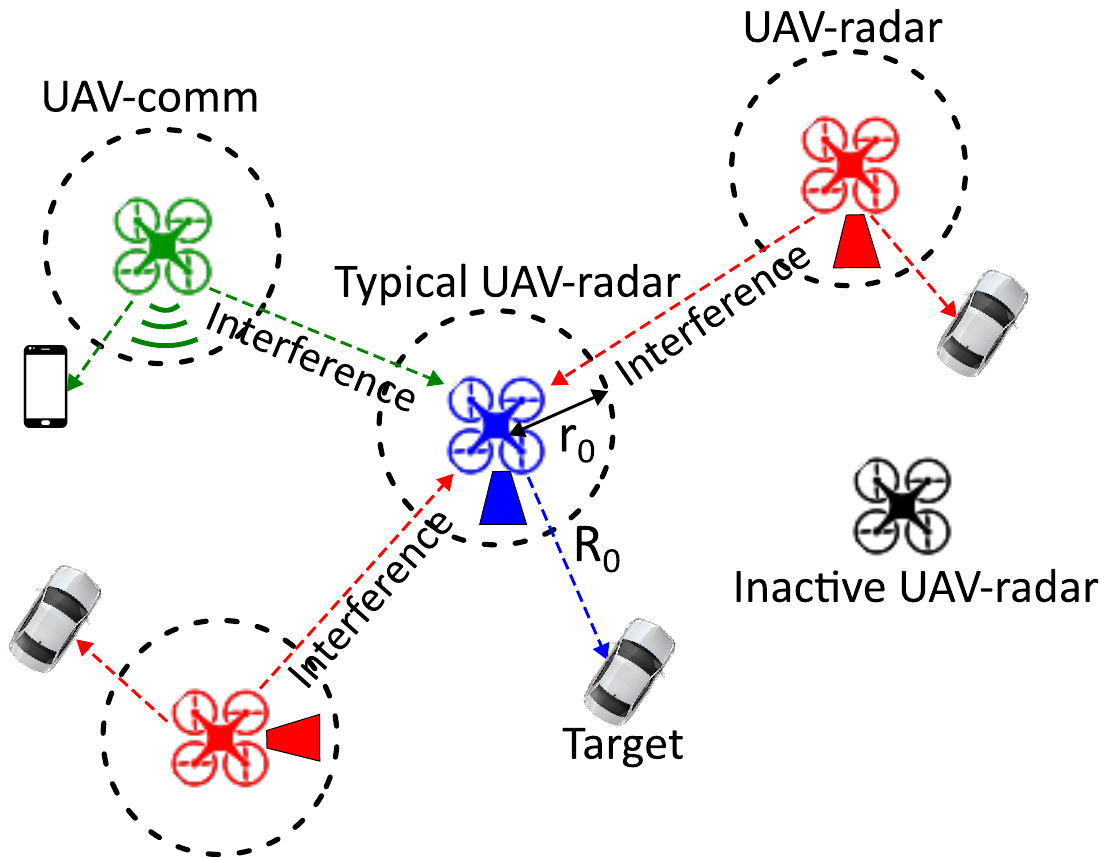}\label{fig:illu_netw_1}}
    
    \subfloat[Communication scenario.]{\includegraphics[width=0.45\textwidth]{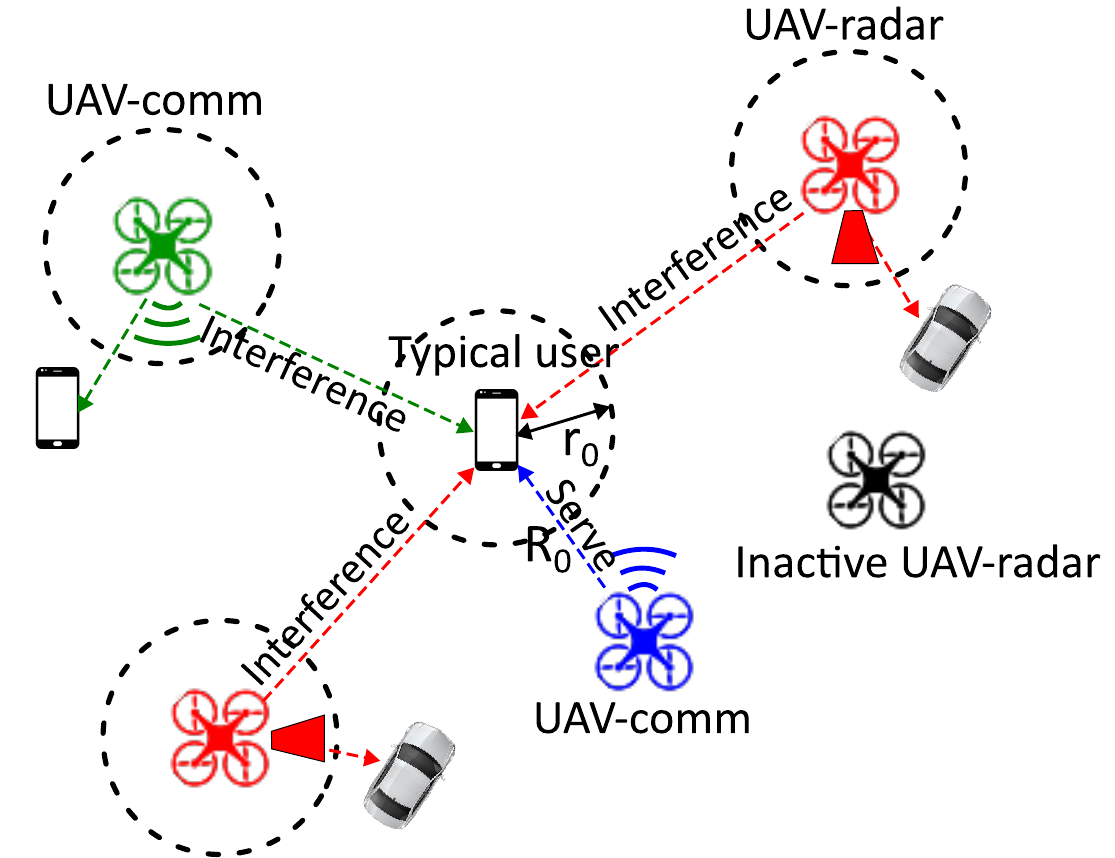}\label{fig:illu_netw_2}}
    \caption{UAV radar and communication network coexistence. A blue UAV is either: (a) a typical UAV-radar that detects a target using radar transmission; or (b) a serving UAV that communicates with a typical user. The green UAV-comms or the red (active) UAV-radars can interfere with radar detection or communication signals. The black (inactive) UAV-radars do not transmit any interference signals. To avoid strong interference, a guard zone is with a radius $r_0$ established between UAVs, and between a UAV and a user.}
    \label{fig:illu_netw}
\end{figure}

 We consider UAV networks where radar nodes and communication nodes coexist. Radar-mounted UAVs (UAV-radars) detect and track a target on the ground by transmitting radar signals and receiving the reflected signals from the target. On the other hand, UAVs  that are equipped with a BS (UAV-comms) communicate with a ground user. We assume that UAV-radars and UAV-comms follow a two-dimensional HPPP independently where the node densities are ${\lambda_{\rm r}}^{\prime}$ and ${\lambda_{\rm d}}^{\prime}$ respectively. All UAVs fly at a fixed identical height $\mathsf{h}_{\rm UAV}$.  Fig.~\ref{fig:illu_netw} describes two different network representations of the radar detection scenario and the communication scenario in the HPPP model. The guard zone with radius $r_0$ is considered between two UAVs, or between a UAV and a user to protect them from potential strong interference. The distance between a UAV-radar and a target in the radar detection scenario and the distance between a UAV-comm and a served user in the communication scenario is $R_0$. UAV-radars are assigned to active UAV-radar by the random spectrum access with the duty cycle $\delta$, and the rest of UAV-radars remain inactive UAV-radars in the networks.

\begin{figure}[t!]
    \centering
    \includegraphics[width=0.48\textwidth]{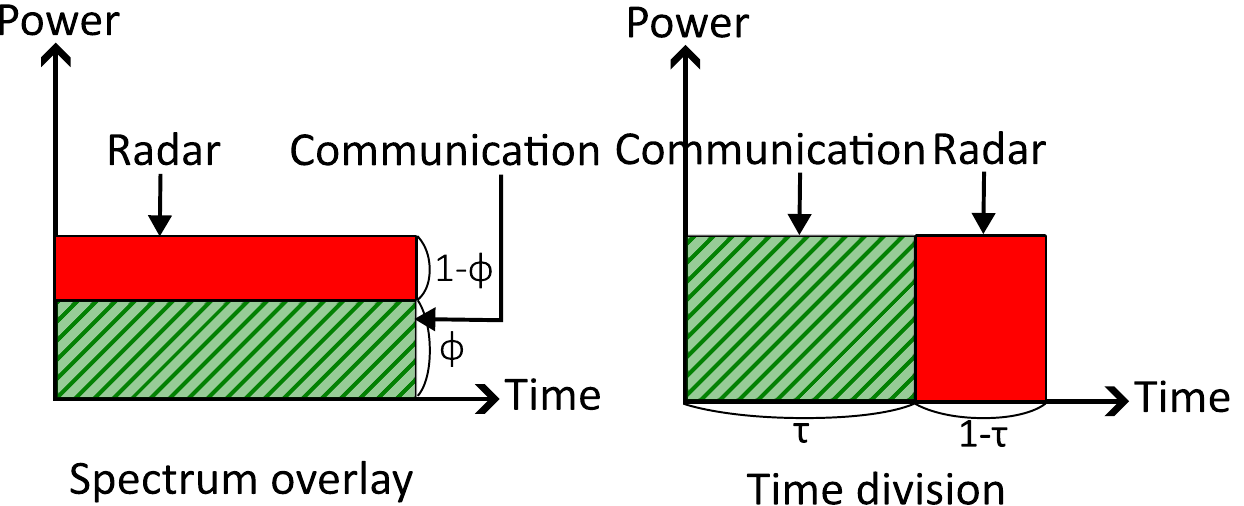}
    \caption{Two different multiple access schemes: SOMA (left) and TDMA (right).}
    \label{fig:illu_ma}
\end{figure}

In the radar and communication coexistence, UAV-radars and UAV-comms need to coordinate the time and spectrum resources for radar and data transmissions. We consider two different multiple access schemes: SOMA where radar signal and data signal share the same spectrum during transmission time, and TDMA where the time for radar and data transmission is scheduled at separate time slots. Fig.~\ref{fig:illu_ma} illustrates different radar and data allocations depending on multiple access schemes. The power allocated to radar and communication is determined by power splitting factor $\phi$ for SOMA, and the time duration assigned to radar and communication is decided by time division factor $\tau$ for TDMA. Note that the interference behavior in this network coexistence  is dependent on the multiple access. In this paper, we focus on the analysis and comparison of SOMA and TDMA on data transmission and radar detection. The key parameters
are summarized in Table II.

\section{Signal Propagation Models in HPPP}\label{Sec:HPPP}
In this section, we describe signal and interference models in HPPP when SOMA and TDMA are adopted respectively. Throughout this paper, we denote SOMA and TDMA as $\rm s.o.$ and $\rm t.d.$ at the superscript.
\subsection{Radar and Data Signal Models}
\subsubsection{SOMA}
We place a typical UAV-radar with the origin $(0,~0,~\mathsf{h}_{\rm UAV})$ and distance from the target at $(x_{\rm t},y_{\rm t},0)$ is $R_0=\sqrt{x_{\rm t}^2+y_{\rm t}^2+(\mathsf{h}_{\rm UAV})^2}$ as in Fig.~\ref{fig:illu_netw_1}. We assume that the height of a UAV-radar $\mathsf{h}_{\rm UAV}$ is sufficiently high so that line-of-sight is secured to detect the target and the free-space path loss model can be considered~\cite{6863654}. The power of the received signal that is reflected back from the target can be expressed as~\cite{7819520},
\begin{align}\label{eq:pr_so}
    \mathsf{P}_{\rm r}^{\rm s.o.}=\left(\frac{(1-\phi)\mathsf{P}_{\rm Tx}\mathsf{G}_{\rm t}}{4\pi R_0^{\alpha}}\right)\left(\frac{\sigma S_{\rm e}}{4\pi R_0^{\alpha}}\right)\mathsf{G}_{\rm p},
\end{align}
where $\mathsf{P}_{\rm Tx}$, $\mathsf{G}_{\rm t}$, $\alpha$ indicate transmit power, Tx antenna gain, and path loss exponent, and $\sigma$, $S_{\rm e}$, $\mathsf{G}_{\rm p}$ denote radar cross-section (RCS) of the target, the effective aperture of radar receiver, and the processing gain. Swerling I model is considered for the RCS and the RCS of the target follows the exponential distribution, $\sigma\sim\frac{1}{\bar{\sigma}}e^{-\frac{\sigma}{\bar{\sigma}}}$~\cite{9119440}. The effective area is given by
\begin{align}
    S_{\rm e}=\frac{\mathsf{G}_{\rm r}c^2}{4\pi f_{\rm c}^2},
\end{align}
where $c$, $f_{\rm c}$ denote the speed of light and the carrier frequency. Then, the power of the reflected back radar signal in \eqref{eq:pr_so} can be rewritten as
\begin{align}\label{eq:pr_so_2}
    \mathsf{P}_{\rm r}^{\rm s.o.}=\frac{(1-\phi)\mathsf{P}_{\rm Tx}\mathsf{G}_{\rm t}\mathsf{G}_{\rm r}\mathsf{G}_{\rm p}c^2\sigma}{(4\pi)^3f_{\rm c}^2R_0^{2\alpha}}.
\end{align}

In the communication scenario as in Fig.~\ref{fig:illu_netw_2}, a typcial user is located at the origin and it receives the signal from the  serving UAV-comm at $(x_{\rm t},y_{\rm t},\mathsf{h}_{\rm UAV})$, which is at a distance of $R_0=\sqrt{x_{\rm t}^2+y_{\rm t}^2+(\mathsf{h}_{\rm UAV})^2}$ from the user. The power of the received signal of the user is given by
\begin{align}
    \mathsf{P}_{\rm d}^{\rm s.o.}=\frac{\phi\mathsf{P}_{\rm Tx}\mathsf{G}_{\rm t}\mathsf{G}_{\rm r}c^2}{(4\pi)^2f_{\rm c}^2R_0^{\alpha}}h_0,
\end{align}
where $h_0\sim\exp(1)$ represents Rayleigh small-scale fading. Note that the allocated power of the radar and the data signals are split by the power splitting factor $\phi$ in SOMA as illustrated in Fig~\ref{fig:illu_ma}.

\subsubsection{TDMA}
The received signal power of the radar signal and the data signal in TDMA are expressed as
\begin{align}
    \mathsf{P}_{\rm r}^{\rm t.d.}=\frac{\mathsf{P}_{\rm Tx}\mathsf{G}_{\rm t}\mathsf{G}_{\rm r}\mathsf{G}_{\rm p}c^2\sigma}{(4\pi)^3f_{\rm c}^2R_0^{2\alpha}},\label{eq:pr_td}\\
    \mathsf{P}_{\rm d}^{\rm t.d.}=\frac{\mathsf{P}_{\rm Tx}\mathsf{G}_{\rm t}\mathsf{G}_{\rm r}c^2}{(4\pi)^2f_{\rm c}^2R_0^{\alpha}}h_0.\label{eq:pd_td}
\end{align}
Since the radar detection and communication are separately conducted in different time slots in TDMA, transmit power is not adjusted as in SOMA.

\subsection{Effective Radar and Communication Node Densities}\label{sec:nd_den}
The PPP in this network introduces the guard zone and it can be modeled by the Matérn hard-core point processes (MHCPP) type-II, which can be further approximated by the HPPP model.
\subsubsection{SOMA}
The approximated effective node density of the radar and the communication from MHCPP type-II can be written as~\cite{5934671}
\begin{align}\label{eq:nd_so}
    \lambda_{\rm r}^{\rm s.o.}=\frac{1-e^{-\bar{\lambda}_{\rm r}^{\rm s.o.}\pi r_0^2}}{\pi r_0^2},~\lambda_{\rm d}^{\rm s.o.}=\frac{1-e^{-\lambda^{\prime}_{\rm d}\pi r_0^2}}{\pi r_0^2},
\end{align}
where ${\bar\lambda}_{\rm r}^{\rm s.o.}=\delta\lambda^{\prime}_{\rm r}$ is the active UAV-radar node density, and $\delta$ is the duty cycle.
\subsubsection{TDMA}
The effective node density of the radar and the communication from MHCPP type-II can be written as
\begin{align}\label{eq:nd_tdma}
    \lambda_{\rm r}^{\rm t.d.}=\frac{1-e^{-\bar{\lambda}^{\rm t.d.}_{\rm r}\pi r_0^2}}{\pi r_0^2},~\lambda_{\rm d}^{\rm t.d.}=\frac{1-e^{-\lambda^{\prime}_{\rm d}\pi r_0^2}}{\pi r_0^2},
\end{align}
where ${\bar\lambda}_{\rm r}^{\rm t.d.}=\frac{\delta}{1-\tau}\lambda^{\prime}_{\rm r}$ is the active UAV-radar node density. Note that the duty cycle $\delta$ in SOMA increases to $\frac{\delta}{1-\tau}$ in TDMA as much as the reduced radar transmission time by $\tau$, since it is assumed that the total number of active UAV-radar nodes during the time period is the same for both SOMA and TDMA. Since the effective node density of UAV-comms for TDMA and SOMA is equal, we merge the notation of the node density as $\lambda_{\rm d}=\lambda_{\rm d}^{\rm s.o.}=\lambda_{\rm d}^{\rm t.d.}$.

\subsection{Interference Models}
In this subsection, we obtain the power of interference coming from nearby active UAV-radars and UAV-comms in the HPPP model.
\subsubsection{SOMA}
Since the radar detection and communication occupy the same spectrum band at the same time, the aggregated interference power from UAV-radars and UAV-comms can be expressed as
\begin{align}\label{eq:int_so}
    \mathsf{I}^{\rm s.o.}&=\underbrace{\sum_{r_i\in\Phi(\lambda_{\rm d})\backslash r_0}\frac{\phi\mathsf{P}_{\rm Tx}\mathsf{G}_{\rm t}\mathsf{G}_{\rm rI}c^2}{(4\pi)^2f_{\rm c}^2r_i^{\alpha_{\rm I}}}h_i}_{\text{interfence from nearby UAV-comms }(\mathsf{I}_1)}\nonumber\\
    &+\underbrace{\sum_{r_j\in\Phi(\lambda_{\rm r}^{\rm s.o.})\backslash r_0}\frac{(1-\phi)\mathsf{P}_{\rm Tx}\mathsf{G}_{\rm t}\mathsf{G}_{\rm rI}c^2}{(4\pi)^2f_{\rm c}^2r_j^{\alpha_{\rm I}}}h_j}_{\text{interfence from nearby UAV-radars }(\mathsf{I}_2)},
\end{align}
where $r\in\Phi(\lambda)\backslash r_0$ means a two-dimensional HPPP with a density $\lambda$ and $r>r_0$, $\mathsf{G}_{\rm rI}$ and $\alpha_{\rm I}$ are Rx antenna gain and path-loss exponent from the interfering signals, respectively, $h_i$, $h_j$ represent small-scale fading from the interference, and $r_i$, $r_j$ are distance between a typical UAV-radar or user and an interferer. Since the radar detection and communication are carried out together in SOMA, the interference power at both the typical UAV-radar (Fig.~\ref{fig:illu_netw_1}) and the typical information receiver (Fig.~\ref{fig:illu_netw_2}) is the same: $\mathsf{I}^{\rm s.o.}=\mathsf{I}^{\rm s.o.}_{\rm d}=\mathsf{I}^{\rm s.o.}_{\rm r}$.
\subsubsection{TDMA}
The interfence comes only from the UAV-radars in the radar detection scenario. Likewise, the interference comes only from UAV-comms in communication scenario. Then the aggregated interference power for each case can be expressed as
\begin{align}
    \mathsf{I}^{\rm t.d.}_{\rm r}&=\sum_{r_i\in\Phi(\lambda_{\rm r}^{\rm t.d.})\backslash r_0}\frac{\mathsf{P}_{\rm Tx}\mathsf{G}_{\rm t}\mathsf{G}_{\rm rI}c^2}{(4\pi)^2f_{\rm c}^2r_i^{\alpha_{\rm I}}}h_i,\label{eq:int_r_td}\\
    \mathsf{I}^{\rm t.d.}_{\rm d}&=\sum_{r_i\in\Phi(\lambda_{\rm d})\backslash r_0}\frac{\mathsf{P}_{\rm Tx}\mathsf{G}_{\rm t}\mathsf{G}_{\rm rI}c^2}{(4\pi)^2f_{\rm c}^2r_i^{\alpha_{\rm I}}}h_i.\label{eq:int_d_td}
\end{align}

\section{Performance Analysis of Radar Detection and Data Communication}\label{Sec:Perform_ana}
In this section, we discuss performance evaluation metrics for two different scenarios. Specifically, we derive the SRP in the radar detection scenario and the TC in the communication scenario.
\subsection{Successful Ranging Probability}
SRP is the probability that a UAV-radar succeeds in detecting the target, which is decided by the signal-to-interference-plus-noise ratio (SINR). SRP is defined by
\begin{align}\label{eq:def_SRP}
    \mathsf{Pr_{s.r.}}(\gamma_{\rm th})=\mathsf{Pr}(\gamma_0>\gamma_{\rm th}),
\end{align}
where $\gamma_0=\frac{\mathsf{P}_{\rm r}}{\mathsf{I}_{\rm r}+\mathsf{N}_0}$ and $\mathsf{N}_0$ are SINR of the received radar signal and the noise power,  respectively. In addition, $\gamma_{\rm th}$ denotes SINR threshold where the target is successfully detected. In what follows, we derive the closed-form expression of SRP in both SOMA and TDMA.
\subsubsection{SOMA}
SRP in \eqref{eq:def_SRP} can be derived from \eqref{eq:pr_so_2}, \eqref{eq:int_so} as follows:
\begin{align}\label{eq:SRP_der_1}
    &\mathsf{Pr_{s.r.}^{\rm s.o.}}(\gamma_{\rm th})=\mathsf{Pr}\left(\frac{\mathsf{P}^{\rm s.o.}_{\rm r}}{\mathsf{I}^{\rm s.o.}+\mathsf{N}_0}>\gamma_{\rm th}\right)\nonumber\\
    &=\mathsf{Pr}\left(\frac{(1-\phi)\mathsf{P}_{\rm Tx}\mathsf{G}_{\rm t}\mathsf{G}_{\rm r}\mathsf{G}_{\rm p}c^2\sigma}{\mathsf{I}^{\rm s.o.}+\mathsf{N}_0}>(4\pi)^3f_{\rm c}^2R_0^{2\alpha}\gamma_{\rm th}\right)\nonumber\\
    &\approx\mathsf{Pr}\left(\sigma>\frac{(4\pi)^3f_{\rm c}^2R_0^{2\alpha}\gamma_{\rm th}\mathsf{I}^{\rm s.o.}}{(1-\phi)\mathsf{P}_{\rm Tx}\mathsf{G}_{\rm t}\mathsf{G}_{\rm r}\mathsf{G}_{\rm p}c^2}\right)\nonumber\\
    &=\int_0^{\infty}\left\{1-F_{\sigma}\left(\frac{(4\pi)^3f_{\rm c}^2R_0^{2\alpha}\gamma_{\rm th}y}{(1-\phi)\mathsf{P}_{\rm Tx}\mathsf{G}_{\rm t}\mathsf{G}_{\rm r}\mathsf{G}_{\rm p}c^2}\right)\right\}f_{\mathsf{I}^{\rm s.o.}}(y){\rm d}y\nonumber\\
    &=\int_0^{\infty}e^{-\frac{(4\pi)^3f_{\rm c}^2R_0^{2\alpha}\gamma_{\rm th}y}{(1-\phi)\mathsf{P}_{\rm Tx}\mathsf{G}_{\rm t}\mathsf{G}_{\rm r}\mathsf{G}_{\rm p}c^2\bar{\sigma}}}f_{\mathsf{I}^{\rm s.o.}}(y){\rm d}y\nonumber\\
    &=\mathcal{L}_{\mathsf{I}^{\rm s.o.}}\left(\frac{(4\pi)^3f_{\rm c}^2R_0^{2\alpha}\gamma_{\rm th}}{(1-\phi)\mathsf{P}_{\rm Tx}\mathsf{G}_{\rm t}\mathsf{G}_{\rm r}\mathsf{G}_{\rm p}c^2\bar{\sigma}}\right)
\end{align}
where the approximation comes from the interference limit regime assumption,  $F_{\sigma}(X)=1-e^{-\frac{X}{\bar{\sigma}}}$ is the cumulative distribution function (CDF) of $\sigma$, and $f_{\mathsf{I}^{\rm s.o.}}(\rm x)$ is the probability density function (PDF) of $\mathsf{I}^{\rm s.o.}$, $\mathcal{L}_{\mathsf{I}^{\rm s.o.}}(z)$ indicates Laplace transform of the PDF of $\mathsf{I}^{\rm s.o.}$. 

$\mathcal{L}_{\mathsf{I}^{\rm s.o.}}(z)$ can be derived as follows. The interference term can be rewritten as $\mathsf{I}^{\rm s.o.}=\mathsf{I}_1 + \mathsf{I}_2$ where $\mathsf{I}_1$, $\mathsf{I}_2$ are the first and the second terms in \eqref{eq:int_so} respectively. Then, we can obtain~\cite{9043513}
\begin{align}\label{eq:SRP_der_2}
    \mathcal{L}_{\mathsf{I}_{1}}(z)&=\exp{\left\{-2\pi\lambda_{\rm d}{\rm A}_1(z)\frac{(z\phi{\rm K}_1)^{\frac{2}{\alpha_{\rm I}}}}{\alpha_{\rm I}}\right\}},\nonumber\\
    \mathcal{L}_{\mathsf{I}_{2}}(z)&=\exp{\left\{-2\pi\lambda_{\rm r}^{\rm s.o.}{\rm A}_2(z)\frac{(z(1-\phi){\rm K}_1)^{\frac{2}{\alpha_{\rm I}}}}{\alpha_{\rm I}}\right\}},
\end{align}
where {\small ${\rm A}_1(z)={\rm B}\left(\frac{2}{\alpha_{\rm I}},1-\frac{2}{\alpha_{\rm I}}\right)-{\rm B}\left(\frac{1}{1+z\phi{\rm K}_1r_0^{-\alpha_{\rm I}}};\frac{2}{\alpha_{\rm I}},1-\frac{2}{\alpha_{\rm I}}\right)$}, {\small ${\rm A}_2(z)={\rm B}\left(\frac{2}{\alpha_{\rm I}},1-\frac{2}{\alpha_{\rm I}}\right)-{\rm B}\left(\frac{1}{1+z(1-\phi){\rm K}_1r_0^{-\alpha_{\rm I}}};\frac{2}{\alpha_{\rm I}},1-\frac{2}{\alpha_{\rm I}}\right)$}, ${\rm K_1}=\frac{\mathsf{P}_{\rm Tx}\mathsf{G}_{\rm t}\mathsf{G}_{\rm rI}c^2}{(4\pi)^2f_{\rm c}^2}$, ${\rm B}(a,b)$ is the beta function, and ${\rm B}(x;a,b)=\int_0^x u^{a-1}(1-u)^{b-1}{\rm d}u$ is the incomplete beta fucntion. Then, we can derive
\begin{align}\label{eq:int_lap_so}
    &\mathcal{L}_{\mathsf{I}^{\rm s.o.}}(z)=\mathcal{L}_{\mathsf{I}_{1}}(z)\mathcal{L}_{\mathsf{I}_{2}}(z)=\nonumber\\
    &\exp{\left\{-2\pi\left(\phi^{\frac{2}{\alpha_{\rm I}}}\lambda_{\rm d}{\rm A}_1(z)+\left(1-\phi\right)^{\frac{2}{\alpha_{\rm I}}}\lambda_{\rm r}^{\rm s.o.}{\rm A}_2(z)\right)\frac{(z{\rm K}_1)^{\frac{2}{\alpha_{\rm I}}}}{\alpha_{\rm I}}\right\}}.
\end{align}
Then, the closed-form expression of SRP in SOMA can be obtained as \eqref{eq:SRP_so} at the top of the next page.
\begin{figure*}[t]
\begin{align}\label{eq:SRP_so}
    &\mathsf{Pr_{s.r.}^{\rm s.o.}}(\gamma_{\rm th})=\exp\left\{-2\pi\left(\left(\frac{\phi}{1-\phi}\right)^{\frac{2}{\alpha_{\rm I}}}\lambda_{\rm d}\underbrace{\left\{{\rm B}\left(\frac{2}{\alpha_{\rm I}},1-\frac{2}{\alpha_{\rm I}}\right)-{\rm B}\left(\frac{1}{1+\left(\frac{\phi4\pi {\rm G}_{\rm rI}R_0^{2\alpha}\gamma_{\rm th}}{(1-\phi)\mathsf{G}_{\rm r}\mathsf{G}_{\rm p}\bar{\sigma}}\right)r_0^{-\alpha_{\rm I}}};\frac{2}{\alpha_{\rm I}},1-\frac{2}{\alpha_{\rm I}}\right)\right\}}_{C_1}+\right.\right.\nonumber\\
    &\left.\left.\lambda_{\rm r}^{\rm s.o.}\underbrace{\left\{{\rm B}\left(\frac{2}{\alpha_{\rm I}},1-\frac{2}{\alpha_{\rm I}}\right)-{\rm B}\left(\frac{1}{1+\left(\frac{4\pi {\rm G}_{\rm rI}R_0^{2\alpha}\gamma_{\rm th}}{\mathsf{G}_{\rm r}\mathsf{G}_{\rm p}\bar{\sigma}}\right)r_0^{-\alpha_{\rm I}}};\frac{2}{\alpha_{\rm I}},1-\frac{2}{\alpha_{\rm I}}\right)\right\}}_{C_2}\right)\frac{\left(\frac{4\pi {\rm G}_{\rm rI}R_0^{2\alpha}\gamma_{\rm th}}{\mathsf{G}_{\rm r}\mathsf{G}_{\rm p}\bar{\sigma}}\right)^{\frac{2}{\alpha_{\rm I}}}}{\alpha_{\rm I}}\right\}
\end{align}
\hrulefill
\begin{align}\label{eq:SRP_td}
    \mathsf{Pr_{s.r.}^{\rm t.d.}}(\gamma_{\rm th})=\exp\left\{-2\pi\lambda_{\rm r}^{\rm t.d.}\underbrace{\left\{{\rm B}\left(\frac{2}{\alpha_{\rm I}},1-\frac{2}{\alpha_{\rm I}}\right)-{\rm B}\left(\frac{1}{1+\left(\frac{4\pi {\rm G}_{\rm rI}R_0^{2\alpha}\gamma_{\rm th}}{\mathsf{G}_{\rm r}\mathsf{G}_{\rm p}\bar{\sigma}}\right)r_0^{-\alpha_{\rm I}}};\frac{2}{\alpha_{\rm I}},1-\frac{2}{\alpha_{\rm I}}\right)\right\}}_{C_2}\frac{\left(\frac{4\pi {\rm G}_{\rm rI}R_0^{2\alpha}\gamma_{\rm th}}{\mathsf{G}_{\rm r}\mathsf{G}_{\rm p}\bar{\sigma}}\right)^{\frac{2}{\alpha_{\rm I}}}}{\alpha_{\rm I}}\right\}
\end{align}
\hrulefill
\begin{align}\label{eq:tc_so_ana}
    &C^{\rm s.o.}={\lambda}_{\rm d}\log(1+\beta_{\rm th})\exp\left\{-2\pi\left(\lambda_{\rm d}\underbrace{\left\{{\rm B}\left(\frac{2}{\alpha_{\rm I}},1-\frac{2}{\alpha_{\rm I}}\right)-{\rm B}\left(\frac{1}{1+\left(\frac{{\rm G}_{\rm rI}R_0^{\alpha}\beta_{\rm th}}{\mathsf{G}_{\rm r}}\right)r_0^{-\alpha_{\rm I}}};\frac{2}{\alpha_{\rm I}},1-\frac{2}{\alpha_{\rm I}}\right)\right\}}_{C_3}+\right.\right.\nonumber\\
    &\left.\left.\left(\frac{1-\phi}{\phi}\right)^{\frac{2}{\alpha_{\rm I}}}\lambda_{\rm r}^{\rm s.o.}\underbrace{\left\{{\rm B}\left(\frac{2}{\alpha_{\rm I}},1-\frac{2}{\alpha_{\rm I}}\right)-{\rm B}\left(\frac{1}{1+\left(\frac{(1-\phi){\rm G}_{\rm rI}R_0^{\alpha}\beta_{\rm th}}{\phi\mathsf{G}_{\rm r}}\right)r_0^{-\alpha_{\rm I}}};\frac{2}{\alpha_{\rm I}},1-\frac{2}{\alpha_{\rm I}}\right)\right\}}_{C_4}\right)\frac{\left(\frac{{\rm G}_{\rm rI}R_0^{\alpha}\beta_{\rm th}}{\mathsf{G}_{\rm r}}\right)^{\frac{2}{\alpha_{\rm I}}}}{\alpha_{\rm I}}\right\}
\end{align}
\hrulefill
\begin{align}\label{eq:tc_td_ana}
    C^{\rm t.d.}=\tau{\lambda}_{\rm d}\log(1+\beta_{\rm th})\exp\left\{-2\pi\lambda_{\rm d}\underbrace{\left\{{\rm B}\left(\frac{2}{\alpha_{\rm I}},1-\frac{2}{\alpha_{\rm I}}\right)-{\rm B}\left(\frac{1}{1+\left(\frac{{\rm G}_{\rm rI}R_0^{\alpha}\beta_{\rm th}}{\mathsf{G}_{\rm r}}\right)r_0^{-\alpha_{\rm I}}};\frac{2}{\alpha_{\rm I}},1-\frac{2}{\alpha_{\rm I}}\right)\right\}}_{C_3}\frac{\left(\frac{{\rm G}_{\rm rI}R_0^{\alpha}\beta_{\rm th}}{\mathsf{G}_{\rm r}}\right)^{\frac{2}{\alpha_{\rm I}}}}{\alpha_{\rm I}}\right\}
\end{align}
\hrulefill
\end{figure*}

\subsubsection{TDMA}
SRP can be derived from \eqref{eq:pr_td}, \eqref{eq:int_r_td} as follow:
\begin{align}\label{eq:SRP_td}
    \mathsf{Pr_{s.r.}^{\rm t.d.}}(\gamma_{\rm th})&=\mathsf{Pr}\left(\frac{\mathsf{P}^{\rm t.d.}_{\rm r}}{\mathsf{I}^{\rm t.d.}_{\rm r}+\mathsf{N}_0}>\gamma_{\rm th}\right),\nonumber\\
    &=\mathcal{L}_{\mathsf{I}^{\rm t.d.}_{\rm r}}\left(\frac{(4\pi)^3f_{\rm c}^2R_0^{2\alpha}\gamma_{\rm th}}{\mathsf{P}_{\rm Tx}\mathsf{G}_{\rm t}\mathsf{G}_{\rm r}\mathsf{G}_{\rm p}c^2\bar{\sigma}}\right),\\
    \mathcal{L}_{\mathsf{I}^{\rm t.d.}_{\rm r}}(z)&=\exp{\left\{-2\pi\lambda_{\rm r}^{\rm t.d.}{\rm A}_3(z)\frac{(z{\rm K}_1)^{\frac{2}{\alpha_{\rm I}}}}{\alpha_{\rm I}}\right\}},
\end{align}
where ${\rm A}_3(z)={\rm B}(\frac{2}{\alpha_{\rm I}},1-\frac{2}{\alpha_{\rm I}})-{\rm B}(\frac{1}{1+z{\rm K}_1r_0^{-\alpha_{\rm I}}};\frac{2}{\alpha_{\rm I}},1-\frac{2}{\alpha_{\rm I}})$. Note that detailed mathematical steps are skipped, since many steps are similar to \eqref{eq:SRP_der_1}, \eqref{eq:SRP_der_2}. Then, the closed-form expression of SRP in TDMA can be expressed as \eqref{eq:SRP_td} at the top of the next page.

\subsection{Transmission Capacity}
TC is defined by the achievable data rate given an outage constraint multiplied by the spatial density and the data transmission time duration~\cite{4712724,8106812}. At first, the outage probability can be expressed as
\begin{align}\label{eq:outage_prob}
    \mathsf{Pr_{out}}(\beta_{\rm th})=\mathsf{Pr}(\beta_0<\beta_{\rm th}),
\end{align}
where $\beta_0=\frac{\mathsf{P}_{\rm d}}{\mathsf{I}_{\rm d}+\mathsf{N}_0}$ is SINR of the received data signal, and $\beta_{\rm th}$ is a target SINR. Then, TC is given  as
\begin{align}
    C^{\rm s.o.}&={\lambda}_{\rm d}(1-\mathsf{Pr_{out}}(\beta_{\rm th}))\log(1+\beta_{\rm th}),\label{eq:tran_cap_so}\\
    C^{\rm t.d.}&=\tau{\lambda}_{\rm d}(1-\mathsf{Pr_{out}}(\beta_{\rm th}))\log(1+\beta_{\rm th}),\label{eq:tran_cap_td}
\end{align}
where $C^{\rm s.o.},~C^{\rm t.d.}$ denote transmission capacity of SOMA and TDMA, respectively. Next, we derive the closed-form expression of TC in both SOMA and TDMA.
\subsubsection{SOMA}
Outage probability in \eqref{eq:outage_prob} can be derived as
\begin{align}\label{eq:op_so_der_1}
    &\mathsf{Pr_{out}^{s.o.}}(\beta_{\rm th})=1-\mathsf{Pr}\left(\frac{\mathsf{P}^{\rm s.o.}_{\rm d}}{\mathsf{I}^{\rm s.o.}+\mathsf{N}_0}>\beta_{\rm th}\right)\nonumber\\
    &=1-\int_0^{\infty}\left\{1-F_{h_0}\left(\frac{(4\pi)^2f^2R_0^{\alpha}\beta_{\rm th}y}{\phi\mathsf{P}_{\rm Tx}\mathsf{G}_{\rm t}\mathsf{G}_{\rm r}c^2}\right)\right\}f_{\mathsf{I}^{\rm s.o.}}(y){\rm d}y\nonumber\\
    &=1-\mathcal{L}_{\mathsf{I}^{\rm s.o.}}\left(\frac{(4\pi)^2f^2R_0^{\alpha}\beta_{\rm th}}{\phi\mathsf{P}_{\rm Tx}\mathsf{G}_{\rm t}\mathsf{G}_{\rm r}c^2}\right),
\end{align}
where $F_{h_0}(X)=1-e^{-X}$. From \eqref{eq:int_lap_so}, \eqref{eq:tran_cap_so}, and \eqref{eq:op_so_der_1}, the closed-from expression of the TC in SOMA is given as \eqref{eq:tc_so_ana} at the top of the next page.
\subsubsection{TDMA}
Outage probability in TDMA can be derived as
\begin{align}\label{eq:op_td_der_1}
    \mathsf{Pr_{out}^{t.d.}}(\beta_{\rm th})&=1-\mathsf{Pr}\left(\frac{\mathsf{P}^{\rm t.d.}_{\rm d}}{\mathsf{I}^{\rm t.d.}_{\rm d}+\mathsf{N}_0}>\beta_{\rm th}\right)\nonumber\\
    &=1-\mathcal{L}_{\mathsf{I}^{\rm t.d.}_{\rm d}}\left(\frac{(4\pi)^2f^2R_0^{\alpha}\beta_{\rm th}}{\mathsf{P}_{\rm Tx}\mathsf{G}_{\rm t}\mathsf{G}_{\rm r}c^2}\right).
\end{align}
The Laplace transform of $\mathsf{I}^{\rm t.d.}_{\rm d}$ can be derived as
\begin{align}\label{eq:lap_int_td_d}
    \mathcal{L}_{\mathsf{I}^{\rm t.d.}_{\rm d}}(z)&=\exp{\left\{-2\pi\lambda_{\rm d}\int_{r_0}^{\infty}\mathbb{E}_{h}\left[1-e^{-z{\rm K}_1hr^{-\alpha_{\rm I}}}\right]r{\rm d}r\right\}}\nonumber\\
    &=\exp{\left\{-2\pi\lambda_{\rm d}{\rm A}_3(z)\frac{(z{\rm K}_1)^{\frac{2}{\alpha_{\rm I}}}}{\alpha_{\rm I}}\right\}}.
\end{align}
From \eqref{eq:tran_cap_so}, \eqref{eq:op_td_der_1}, \eqref{eq:lap_int_td_d}, the closed-form expression of the TC in TDMA can be derived as \eqref{eq:tc_td_ana} at the top of the page.

\section{Network Design Strategy for Successful Ranging Probability}\label{Sec:SRP_ana}
In this section, we discuss how network parameters such as node densities, radius of guard zone, power splitting factor, and time division factor are determined from the analysis in Section~\ref{Sec:Perform_ana} for a given SRP constraint.
\subsection{Node Densities}\label{sec:SRP_nd}
The density of the node in the networks affects the power of the interference signal. Specifically, as the UAV-radar node density $\lambda_{\rm r}^{\prime}$ increases, the interference power at the typical UAV-radar increases and therefore, the SINR of the received radar signal decreases, which results in the lower SRP. In SOMA, SINR is also affected by the UAV-comm node density $\lambda_{\rm d}^{\prime}$ due to the simultaneous transmission of data and radar signals. Therefore, one can be interested in finding the maximum node density given a target SRP ($\mathsf{\bar{Pr}_{s.r.}}$) and SINR threshold $\gamma_{\rm th}$ of the SRP.
\subsubsection{SOMA}
When SINR threshold $\gamma_{\rm th}$ and the target SRP are given, we can rearrange \eqref{eq:SRP_so} such that only the terms that are related to the node densities $\lambda_{\rm d}$ and $\lambda_{\rm r}^{\rm s.o.}$ are placed to the left side of the equation. Then, we obtain inequality as follows: 
\begin{align}\label{eq:SRP_ineq}
    \left(\frac{\phi}{1-\phi}\right)^{\frac{2}{\alpha_{\rm I}}}C_1\lambda_{\rm d} + C_2\lambda_{\rm r}^{\rm s.o.}\leq\frac{-\log\mathsf{\bar{Pr}_{s.r.}^{\rm s.o.}}\alpha_{\rm I}}{2\pi\left(\frac{4\pi {\rm G}_{\rm rI}R_0^{2\alpha}\gamma_{\rm th}}{\mathsf{G}_{\rm r}\mathsf{G}_{\rm p}\bar{\sigma}}\right)^{\frac{2}{\alpha_{\rm I}}}},
\end{align}
where $C_1$ and $C_2$ are indicated in \eqref{eq:SRP_so}. If we assume a condition $\lambda_{\rm d}=\lambda_{\rm r}^{\rm s.o.}$, the above inequality can be rewritten as
\begin{align}\label{eq:SRP_ndmax}
    \lambda_{\rm d}=\lambda_{\rm r}^{\rm s.o.}\leq\frac{-\log\mathsf{\bar{Pr}_{s.r.}^{\rm s.o.}}\alpha_{\rm I}}{2\pi\left(\frac{4\pi {\rm G}_{\rm rI}R_0^{2\alpha}\gamma_{\rm th}}{\mathsf{G}_{\rm r}\mathsf{G}_{\rm p}\bar{\sigma}}\right)^{\frac{2}{\alpha_{\rm I}}}\left(\left(\frac{\phi}{1-\phi}\right)^{\frac{2}{\alpha_{\rm I}}}C_1 + C_2\right)}.
\end{align}
The maximum node densities $\lambda_{\rm d}^{\star}$ and $\lambda_{\rm r}^{\rm s.o.\star}$ can be obtained when \eqref{eq:SRP_ndmax} goes to equality. Note that this analysis can be easily extended to the condition that $\lambda_{\rm d}$ and $\lambda_{\rm r}^{\rm s.o.}$ are given by the different ratio ($\lambda_{\rm d}\propto\lambda_{\rm r}^{\rm s.o.}$) to find the maximum node densities.
\subsubsection{TDMA}
In the same manner of obtaining \eqref{eq:SRP_ndmax} for SOMA, the maximum node densities of the UAV-radar $\lambda_{\rm r}^{\rm t.d.\star}$ with the given target SRP and SINR threshold can be expressed from \eqref{eq:SRP_td} as
\begin{align}
    \lambda_{\rm r}^{\rm t.d.\star}=\frac{-\log\mathsf{\bar{Pr}_{s.r.}^{\rm s.o.}}\alpha_{\rm I}}{2\pi\left(\frac{4\pi {\rm G}_{\rm rI}R_0^{2\alpha}\gamma_{\rm th}}{\mathsf{G}_{\rm r}\mathsf{G}_{\rm p}\bar{\sigma}}\right)^{\frac{2}{\alpha_{\rm I}}}C_2}.
\end{align}

\subsection{Radius of Guard Zone}\label{sec:SRP_radius}
Guard zone constrains the minimum distance between nodes to avoid strong interference. As the minimum distance increases, the power of the interference decreases. This implies that  SRP is reduced as  radius of guard zone $r_0$ increases. When we design networks with target SRP ($\mathsf{\bar{Pr}_{s.r.}}$) and the SINR threshold $\gamma_{\rm th}$, the minimum radius of the guard zone $r_0$ that satisfies the target performance can be obtain by solving \eqref{eq:SRP_so} in SOMA and \eqref{eq:SRP_td} in TDMA for $r_0$.

\subsection{Power Splitting Factor $\phi$ in SOMA}\label{sec:SRP_phi}
Power splitting factor $\phi$ determines the transmit power ratio between UAV-comms and UAV-radars in SOMA where the radar signal power proportionally decreases as $\phi$ increases. From the closed-form expression of the SRP in \eqref{eq:SRP_so}, the terms that are affected by $\phi$ are $\left(\frac{\phi}{1-\phi}\right)^{\frac{2}{\alpha_{\rm I}}}$ and ${\rm B}\left(\frac{1}{1+\left(\frac{\phi4\pi {\rm G}_{\rm rI}R_0^{2\alpha}\gamma_{\rm th}}{(1-\phi)\mathsf{G}_{\rm r}\mathsf{G}_{\rm p}\bar{\sigma}}\right)r_0^{-\alpha_{\rm I}}};\frac{2}{\alpha_{\rm I}},1-\frac{2}{\alpha_{\rm I}}\right)$. Since $\frac{\phi}{1-\phi}$ and incomplete beta function are monotonic increasing functions, it is easily proved that  SRP is decreasing function with respect to the power splitting factor $\phi$. This can be intuitively interpreted as higher transmit power of UAV-comms increasing the power of the interference signal. 

\newtheorem{Proposition}{Proposition}
\begin{Proposition}
    When $0\leq\phi<0.5$, the impact of the UAV-comm node density $\lambda_{\rm d}$ on SRP is less than the UAV-radar node density $\lambda_{\rm r}^{\rm s.o.}$. When $\phi=0.5$, the impact of the communication node density $\lambda_{\rm d}$ on SRP is equal to the radar node density $\lambda_{\rm r}^{\rm s.o.}$, while when $0.5<\phi\leq 1$, the impact of the communication node density $\lambda_{\rm d}$ on SRP is greater than the radar node density $\lambda_{\rm r}^{\rm s.o.}$.
\end{Proposition}
\begin{IEEEproof}
From \eqref{eq:SRP_so}, we can observe that varying $\phi$ only affects the node density of UAV-comm $\lambda_{\rm d}$ term, not the node density of UAV-radar $\lambda_{\rm r}^{\rm s.o.}$ term. Then, when $\phi=0.5$, $\frac{\phi}{1-\phi}$ becomes 1, which leads to the result that the impact of $\lambda_{\rm d}$ becomes the same as the impact of $\lambda_{\rm r}^{\rm s.o.}$. On the other hand, when $\phi$ is greater than 0.5, $\frac{\phi}{1-\phi}$ becomes greater  than 1 as well, which makes the multiplying term by $\lambda_{\rm d}$ becomes greater than the multiplying term by $\lambda_{\rm r}^{\rm s.o.}$. In the same way, when $\phi$ is less than 0.5, the multiplying term by $\lambda_{\rm d}$ becomes less than the multiplying term by $\lambda_{\rm r}^{\rm s.o.}$.
\end{IEEEproof}

\textit{Propostion 1} implies that  SRP is affected by the ratio between the node density of UAV-comm ($\lambda_{\rm d}$) and the UAV-radar ($\lambda_{\rm r}^{\rm s.o.}$) and when $\phi$ is given, a different ratio of UAV-comm and UAV-radar node density can improve SRP, which is observed in Fig.~\ref{fig:SRP_phi_ratio} of Section~\ref{Sec:Results}.

\subsection{Time Division Factor $\tau$ in TDMA}\label{sec:SRP_tau}
As we mention in Section~\ref{sec:nd_den}, the increase in $\tau$ reduces radar transmission time and increase the duty cycle, which results in higher node density of the active UAV-radar $\bar{\lambda}^{\rm t.d.}_{\rm r}$.  The effective UAV-radar node density $\lambda_{\rm r}^{\rm t.d.}$ in the HPPP approximation is proportionally increased by $\bar{\lambda}^{\rm t.d.}_{\rm r}$ in \eqref{eq:nd_tdma}. 

\section{Network Design Strategy for Transmission Capacity}\label{Sec:TC_ana}
In this section, we analyze  TC depending on network design parameters. We find the node densities that maximize the TC and we investigate the impact of the radius of  guard zone. We also investigate the effect of the power splitting factor and the time division factor on the TC.

\subsection{Node Densities}\label{sec:TC_nd}
As the node density of the UAV-comm $\lambda_{\rm d}$ increases,  SINR is decreased by the larger number of interferers but the higher node density can increase the capacity of the unit area. Because of this trade-off, we can find the maximum node density $\lambda_{\rm d}$ that maximizes  TC.
\subsubsection{SOMA}
When target SINR $\beta_{\rm th}$, the UAV-radar node density $\lambda_{\rm r}^{\rm s.o.}$,  radius of  guard zone $r_0$, and the power splitting factor $\phi$ are given, we can find the $\lambda_{\rm d}$ that maximizes the TC from \eqref{eq:tc_so_ana}. The term in \eqref{eq:tc_so_ana} that is affected by $\lambda_{\rm d}$ are written as
\begin{align}\label{eq:nd_so_ana_1}
    D_1 = {\lambda}_{\rm d}\log(1+\beta_{\rm th})\exp\left(-2\pi{\lambda}_{\rm d}C_3\frac{\left(\frac{{\rm G}_{\rm rI}R_0^{\alpha}\beta_{\rm th}}{\mathsf{G}_{\rm r}}\right)^{\frac{2}{\alpha_{\rm I}}}}{\alpha_{\rm I}}\right),
\end{align}
where $C_3$ is indicated in \eqref{eq:tc_so_ana}. From \eqref{eq:nd_so_ana_1}, the first and the second derivative of transmission capacity with respective to ${\lambda}_{\rm d}$ can be expressed as
\begin{align}
    (C^{\rm s.o.})^{'}&=\log(1+\beta_{\rm th})\exp\left(-2\pi{\lambda}_{\rm d}C_3^{\prime}\right)\left(1-2\pi{\lambda}_{\rm d}C_3^{\prime}\right),\\
    (C^{\rm s.o.})^{''}&=4\pi C_3^{\prime}\log(1+\beta_{\rm th})\exp\left(-2\pi{\lambda}_{\rm d}C_3^{\prime}\right)\left(\pi{\lambda}_{\rm d}C_3^{\prime}-1\right),
\end{align}
where $C_3^{\prime}=C_3\frac{\left(\frac{{\rm G}_{\rm rI}R_0^{\alpha}\beta_{\rm th}}{\mathsf{G}_{\rm r}}\right)^{\frac{2}{\alpha_{\rm I}}}}{\alpha_{\rm I}}$. Then,  transmission capacity is maximized at 
\begin{align}
    {\lambda}_{\rm d}^{\star}=\frac{1}{2\pi C_3^{\prime}}\quad\quad\text{(SOMA)}.
\end{align}
In SOMA, the node density of UAV-radar $\lambda_{\rm r}^{\rm s.o.}$ also increases the power of interference, which reduces the TC. In $\eqref{eq:tc_so_ana}$, the terms that include the radar node density $\lambda_{\rm r}^{\rm s.o.}$ are given as
\begin{align}
    D_2 = {\lambda}_{\rm d}\log(1+\beta_{\rm th})\exp\left(-2\pi\left(\frac{1-\phi}{\phi}\right)^{\frac{2}{\alpha_{\rm I}}}{\lambda}_{\rm r}^{\rm s.o.}C_4^{\prime}\right),
\end{align}
where $C_4^{\prime}=C_4\frac{\left(\frac{{\rm G}_{\rm rI}R_0^{\alpha}\beta_{\rm th}}{\mathsf{G}_{\rm r}}\right)^{\frac{2}{\alpha_{\rm I}}}}{\alpha_{\rm I}}$. From the above equation, it can be found that  TC is a decreasing function of the ${\lambda}_{\rm r}^{\rm s.o.}$.
\subsubsection{TDMA}
Similarly to SOMA, we can optimize the UAV-comm node density $\lambda_{\rm d}$ in TDMA. From $\eqref{eq:tc_td_ana}$, the TC can be rewritten as
\begin{align}
    C^{\rm t.d.} = \tau{\lambda}_{\rm d}\log(1+\beta_{\rm th})\exp\left(-2\pi{\lambda}_{\rm d}C_3^{\prime}\right).
\end{align}
Then, the optimal UAV-comm node density that maximizes  TC can be derived as
\begin{align}
    {\lambda}_{\rm d}^{\star}=\frac{1}{2\pi C_3^{\prime}}\quad\quad\text{(TDMA)}.
\end{align}
In addition, the TC in TDMA is not affected by $\lambda_{\rm r}^{\rm t.d.}$.
\begin{Remark}
    From the above analysis,  TC is maximized at ${\lambda}_{\rm d}^{\star}=\frac{1}{2\pi C_3^{\prime}}$ for both SOMA and TDMA. On the other hand, the TC in SOMA decreases as $\lambda_{\rm r}^{\rm s.o.}$ increases, while  TC in TDMA is independent of $\lambda_{\rm r}^{\rm t.d.}$.
\end{Remark}

\subsection{Radius of Guard Zone}\label{sec:TC_radius}
Guard zone improves  SINR and it reduces the effective node density $\lambda_{\rm d}$ from \eqref{eq:nd_so}. Therefore, as  radius of guard zone, $r_0$, increases TC would be either improved by  higher SINR or degraded by the lower node density. Since it is mathematically intractable to obtain the first and the second derivatives of  TC with respect to $r_0$ in \eqref{eq:tc_so_ana} and \eqref{eq:tc_td_ana}, we observe the effect of $r_0$ by simulations. From simulation results in Fig.~\ref{fig:TC_SINR}, it is observed that the maximum TC decreases as the $r_0$ increases from 5~m to 25~m, which implies that the TC is a decreasing function of $r_0$ in a typical parameter setup.

\subsection{Power Splitting Factor $\phi$ and Time Division Factor $\tau$}\label{sec:TC_phi_tau}
In SOMA, TC is improved as $\phi$ increases since the transmit power of UAV-comm becomes higher, which improves  SINR. The terms in \eqref{eq:tc_so_ana} that are affected by $\phi$ are $\left(\frac{1-\phi}{\phi}\right)^{\frac{2}{\alpha_{\rm I}}}$ and ${\rm B}\left(\frac{1}{1+\left(\frac{(1-\phi){\rm G}_{\rm rI}R_0^{\alpha}\beta_{\rm th}}{\phi\mathsf{G}_{\rm r}}\right)r_0^{-\alpha_{\rm I}}};\frac{2}{\alpha_{\rm I}},1-\frac{2}{\alpha_{\rm I}}\right)$. Since $\frac{1-\phi}{\phi}$ is a decreasing function and an incomplete beta function is a monotonic increasing function, it is easily proved that TC is an increasing function in terms of $\phi$.

In TDMA, $\tau$ decides the time duration of the data transmission, and larger $\tau$ increases TC. From \eqref{eq:tc_td_ana}, it is observed that TC is linearly increasing with respect to $\tau$.

\section{Performance Comparison of SOMA and TDMA}\label{Sec:SOMA_TDMA}
We compare the performance of  SRP and  transmission capacity between two different multiple access strategies to give intuition in the design of UAV radar sensing and communication network
coexistence. We consider two different scenarios when $\phi = \tau = 0.5$: case 1 and case 2. In case 1, we analyze the condition where the node density of UAV-comms and active UAV-radars is equal, and we compare SOMA with TDMA by SRP and TC. In case 2, we analyze the condition that the node density of UAV-radars is greater than that of UAV-comms, and we find the condition that both SRP and TC of SOMA are higher than those of TDMA.

\subsection{Case 1: $\lambda_{\rm d}^{\prime}=\bar{\lambda}_{\rm r}\neq0$}\label{sec:case 1}
We first analyze a special case that $\lambda_{\rm d}^{\prime}=\bar{\lambda}_{\rm r}\neq0$ and $\phi=\tau=0.5$ where the active UAV-radar and the UAV-comm node density are equal and the resources allocation of the data transmission and the radar detection are the same. In this condition,  SRP of  SOMA and  TDMA can be rewritten from \eqref{eq:SRP_so} and \eqref{eq:SRP_td} as
\begin{align}\label{eq:SRP_com}
    \mathsf{Pr_{s.r.}^{\rm s.o.}}(\gamma_{\rm th})&=\exp\left\{-4\pi\lambda_{\rm r}^{\rm s.o.}C_2\frac{\left(\frac{4\pi {\rm G}_{\rm rI}R_0^{2\alpha}\gamma_{\rm th}}{\mathsf{G}_{\rm r}\mathsf{G}_{\rm p}\bar{\sigma}}\right)^{\frac{2}{\alpha_{\rm I}}}}{\alpha_{\rm I}}\right\},\nonumber\\
    \mathsf{Pr_{s.r.}^{\rm t.d.}}(\gamma_{\rm th})&=\exp\left\{-2\pi\lambda_{\rm r}^{\rm t.d.}C_2\frac{\left(\frac{4\pi {\rm G}_{\rm rI}R_0^{2\alpha}\gamma_{\rm th}}{\mathsf{G}_{\rm r}\mathsf{G}_{\rm p}\bar{\sigma}}\right)^{\frac{2}{\alpha_{\rm I}}}}{\alpha_{\rm I}}\right\}.
\end{align}
Then, we can derive the following proposition.
\begin{Proposition}
    In case 1,  SRP of  TDMA is always greater than  SOMA: $\mathsf{Pr_{s.r.}^{\rm s.o.}}(\gamma_{\rm th})<\mathsf{Pr_{s.r.}^{\rm t.d.}}(\gamma_{\rm th})$.
\end{Proposition}
\begin{IEEEproof}
    From \eqref{eq:SRP_com}, proposition 2 is proved if $2\lambda_{\rm r}^{\rm s.o.}>\lambda_{\rm r}^{\rm t.d.}$ holds. From \eqref{eq:nd_so}, \eqref{eq:nd_tdma}, the statement can be derived as
    \begin{align}
        &2-2e^{-\bar{\lambda}_{\rm r}^{\rm s.o.}\pi r_0^2}>1-e^{-\bar{\lambda}_{\rm r}^{\rm t.d.}\pi r_0^2}\nonumber\\
        \to&2-2e^{-\lambda_{\rm r}^{\prime}\delta\pi r_0^2}>1-e^{-2\lambda_{\rm r}^{\prime}\delta\pi r_0^2}\nonumber\\
        \to&\left(e^{-\lambda_{\rm r}^{\prime}\delta\pi r_0^2}\right)^2-2e^{-\lambda_{\rm r}^{\prime}\delta\pi r_0^2}+1>0\nonumber\\
        \to&\left(e^{-\lambda_{\rm r}^{\prime}\delta\pi r_0^2}-1\right)^2>0.
    \end{align}
\end{IEEEproof}
Next, in this special case,  TC can be rewritten from \eqref{eq:tc_so_ana}, \eqref{eq:tc_td_ana} as
\begin{align}\label{eq:tc_com}
    C^{\rm s.o.}&={\lambda}_{\rm d}\log(1+\beta_{\rm th})\exp\left\{-4\pi\lambda_{\rm d}C_3^{\prime}\right\},\nonumber\\
    C^{\rm t.d.}&=\frac{1}{2}{\lambda}_{\rm d}\log(1+\beta_{\rm th})\exp\left\{-2\pi\lambda_{\rm d}C_3^{\prime}\right\}.
\end{align}
Then, we can derive the following proposition.
\begin{Proposition}
    In case 1,  transmission capacity of  SOMA is greater than  TDMA, when  outage probability $\mathsf{Pr_{out}^{t.d.}}(\beta_{\rm th})<\frac{1}{2}$, $\mathsf{Pr_{out}^{s.o.}}(\beta_{\rm th})<\frac{3}{4}$.
\end{Proposition}
\begin{IEEEproof}
    From \eqref{eq:tc_com}, $C^{\rm s.o.}>C^{\rm t.d.}$, if the following inequality holds:
    \begin{align}
        &{\lambda}_{\rm d}\log(1+\beta_{\rm th})\left(\exp\left\{-2\pi\lambda_{\rm d}C_3^{\prime}\right\}\right)^2\nonumber\\
        &>\frac{1}{2}{\lambda}_{\rm d}\log(1+\beta_{\rm th})\exp\left\{-2\pi\lambda_{\rm d}C_3^{\prime}\right\}\nonumber\\
        \to&\quad\quad\exp\left\{-2\pi\lambda_{\rm d}C_3^{\prime}\right\}>\frac{1}{2},\nonumber\\
        \to&\quad\quad1-\mathsf{Pr_{out}^{t.d.}}(\beta_{\rm th})>\frac{1}{2},\nonumber\\
        \to&\quad\quad\mathsf{Pr_{out}^{t.d.}}(\beta_{\rm th})<\frac{1}{2},\nonumber\\
        \to&\quad\quad\mathsf{Pr_{out}^{s.o.}}(\beta_{\rm th})=1-\left(\exp\left\{-2\pi\lambda_{\rm d}C_3^{\prime}\right\}\right)^2<\frac{3}{4}.
    \end{align}
\end{IEEEproof}
Note that the condition that  outage probability is greater than $\frac{3}{4}$ is a generally desirable condition. Therefore, in case~1  ($\lambda_{\rm d}^{\prime}=\bar{\lambda}_{\rm r}\neq0$ and $\phi=\tau=0.5$), TDMA outperforms  SOMA for  SRP, but SOMA is better than TDMA for  TC. 

\subsection{Case 2: $\lambda_{\rm d}^{\prime}<\bar{\lambda}_{\rm r}$}\label{sec:case 2}
We can also analyze another special case where $\lambda_{\rm d}^{\prime}<\bar{\lambda}_{\rm r}$, $\phi=\tau=0.5$, and an additional condition that the UAV-radar node density $\lambda_{\rm r}^{\prime}$ is sufficiently small. Then, the effective UAV-radar node densities $\lambda_{\rm r}^{\rm s.o.}$, $\lambda_{\rm r}^{\rm t.d.}$ in \eqref{eq:nd_so}, \eqref{eq:nd_tdma} can be approximated by the first order Taylor expansion at $\lambda_{\rm r}^{\prime}=0$ as $\lambda_{\rm r}^{\rm s.o.}\approx\delta\lambda_{\rm r}^{\prime}$ and $\lambda_{\rm r}^{\rm t.d.}\approx 2\delta\lambda_{\rm r}^{\prime}$. In this condition, we can have the following proposition.
\begin{Proposition}
    In case 2 where $\lambda_{\rm d}^{\prime}<\bar{\lambda}_{\rm r}$, the radar node density $\lambda_{\rm r}^{\prime}$ is sufficient small, and $\phi=\tau=0.5$,  SRP of  SOMA is greater than  TDMA.
\end{Proposition}
\begin{IEEEproof}
    From \eqref{eq:SRP_so} and \eqref{eq:SRP_td}, we can obtain  SRP in case~2 as follows:
    \begin{align}\label{eq:SRP_com2}
        \mathsf{Pr_{s.r.}^{\rm s.o.}}(\gamma_{\rm th})&=\exp\left\{-2\pi(\lambda_{\rm d} + \delta\lambda_{\rm r}^{\prime})C_2\frac{\left(\frac{4\pi {\rm G}_{\rm rI}R_0^{2\alpha}\gamma_{\rm th}}{\mathsf{G}_{\rm r}\mathsf{G}_{\rm p}\bar{\sigma}}\right)^{\frac{2}{\alpha_{\rm I}}}}{\alpha_{\rm I}}\right\},\nonumber\\
        \mathsf{Pr_{s.r.}^{\rm t.d.}}(\gamma_{\rm th})&=\exp\left\{-2\pi(2\delta\lambda_{\rm r}^{\prime})C_2\frac{\left(\frac{4\pi {\rm G}_{\rm rI}R_0^{2\alpha}\gamma_{\rm th}}{\mathsf{G}_{\rm r}\mathsf{G}_{\rm p}\bar{\sigma}}\right)^{\frac{2}{\alpha_{\rm I}}}}{\alpha_{\rm I}}\right\}.
    \end{align}
Then, we can easily prove that $\mathsf{Pr_{s.r.}^{\rm s.o.}}>\mathsf{Pr_{s.r.}^{\rm t.d.}}(\gamma_{\rm th})$ if  $\lambda_{\rm d}^{\prime}<\delta\lambda_{\rm r}^{\prime}=\bar{\lambda}_{\rm r}$.
\end{IEEEproof}
Next, we can also obtain the following proposition regarding  TC.
\begin{Proposition}
    In the case that $\phi=\tau=0.5$,  TC of  SOMA is greater than TDMA, when $\mathsf{Pr}(\beta_1<\beta_{\rm th})<\frac{1}{2}$, where
    $\beta_1=\frac{\mathsf{P}^{\rm s.o.}_{\rm r}}{\mathsf{I}_1}$ denotes  SIR in SOMA considering the interference only comes from the active UAV-radar nodes ( $\mathsf{I}_1$  in \eqref{eq:int_so}).
\end{Proposition}
\begin{IEEEproof}
    In the case that $\phi=\tau=0.5$,  transmission capacity can be rewritten from \eqref{eq:tc_so_ana}, \eqref{eq:tc_td_ana} as
\begin{align}\label{eq:tc_com2}
    C^{\rm s.o.}&={\lambda}_{\rm d}\log(1+\beta_{\rm th})\exp\left\{-2\pi\lambda_{\rm d}C_3^{\prime}\right\}\exp\left\{-2\pi\lambda_{\rm r}^{\rm s.o.}C_3^{\prime}\right\},\nonumber\\
    C^{\rm t.d.}&=\frac{1}{2}{\lambda}_{\rm d}\log(1+\beta_{\rm th})\exp\left\{-2\pi\lambda_{\rm d}C_3^{\prime}\right\}.
\end{align}
Then, $C^{\rm s.o.}>C^{\rm t.d.}$ holds, when
\begin{align}
    &1-\exp\left\{-2\pi\lambda_{\rm r}^{\rm s.o.}C_3^{\prime}\right\}<\frac{1}{2}\nonumber\\
    \to~&\mathsf{Pr}(\beta_1<\beta_{\rm th})<\frac{1}{2},
\end{align}
where $\mathsf{I}_{1}=\sum_{r_j\in\Phi(\lambda_{\rm r}^{\rm s.o.})\backslash r_0}\frac{\frac{1}{2}\mathsf{P}_{\rm Tx}\mathsf{G}_{\rm t}\mathsf{G}_{\rm rI}c^2}{(4\pi)^2f_{\rm c}^2r_j^{\alpha_{\rm I}}}h_j$.
\end{IEEEproof}
\begin{Remark}
    From \textit{Proposition 4} and \textit{Proposition 5},  SOMA can outperform  TDMA in both  SRP and  TC, if the conditions in \textit{Proposition 4} and \textit{Proposition 5} are satisfied. This implies that the active UAV-radar node density is greater than the UAV-comm node density ($\lambda_{\rm d}^{\prime}<\bar{\lambda}_{\rm r}$) while the outage probability considering the interference only from the active UAV-radars is less than 0.5. Moreover, when the first condition holds, the second condition is generally desirable since the target outage probability is mostly less than 0.5 and the interference coming from the UAV-comms is smaller than the active UAV-radars.
\end{Remark}

\begin{table}[!t]
\renewcommand{\arraystretch}{1.1}
\caption{Parameter settings for UAV radar and communication network coexistence  analysis.}
\label{table:settings}
\centering
\begin{tabular}{lc}
\hline
Parameter & Value \\
\hline\hline
Transmit power ($\mathsf{P}_\mathsf{Tx}$) & $20$~dBm\\
Transmitter antenna gain ($\mathsf{G}_{\rm t}$) & 10~dBi\\
Receiver antenna gain ($\mathsf{G}_{\rm r}$) & 10~dBi\\
Receiver antenna gain from the interference ($\mathsf{G}_{\rm rI}$) & -10~dBi\\
Target distance ($R_0$) & 50~m\\
Average RCS ($\bar{\sigma}$) & 30~dBsm\\
Path-loss exponent ($\alpha$) & 2.0\\
Path-loss exponent from the interference ($\alpha_{\rm I}$) & 2.5\\
Processing gain ($\mathsf{G}_{\rm p}$) & 10~dBi\\
Duty cycle ($\delta$) & 0.1\\
Carrier frequency ($f_{\rm c}$) & $35$~GHz \\\hline
\hline
\end{tabular}
\vspace{-0.2in}
\end{table}

\begin{figure}[t!]
	\centering
	\subfloat[SINR threshold vs. SRP (case 1)]{\includegraphics[width=0.48\textwidth]{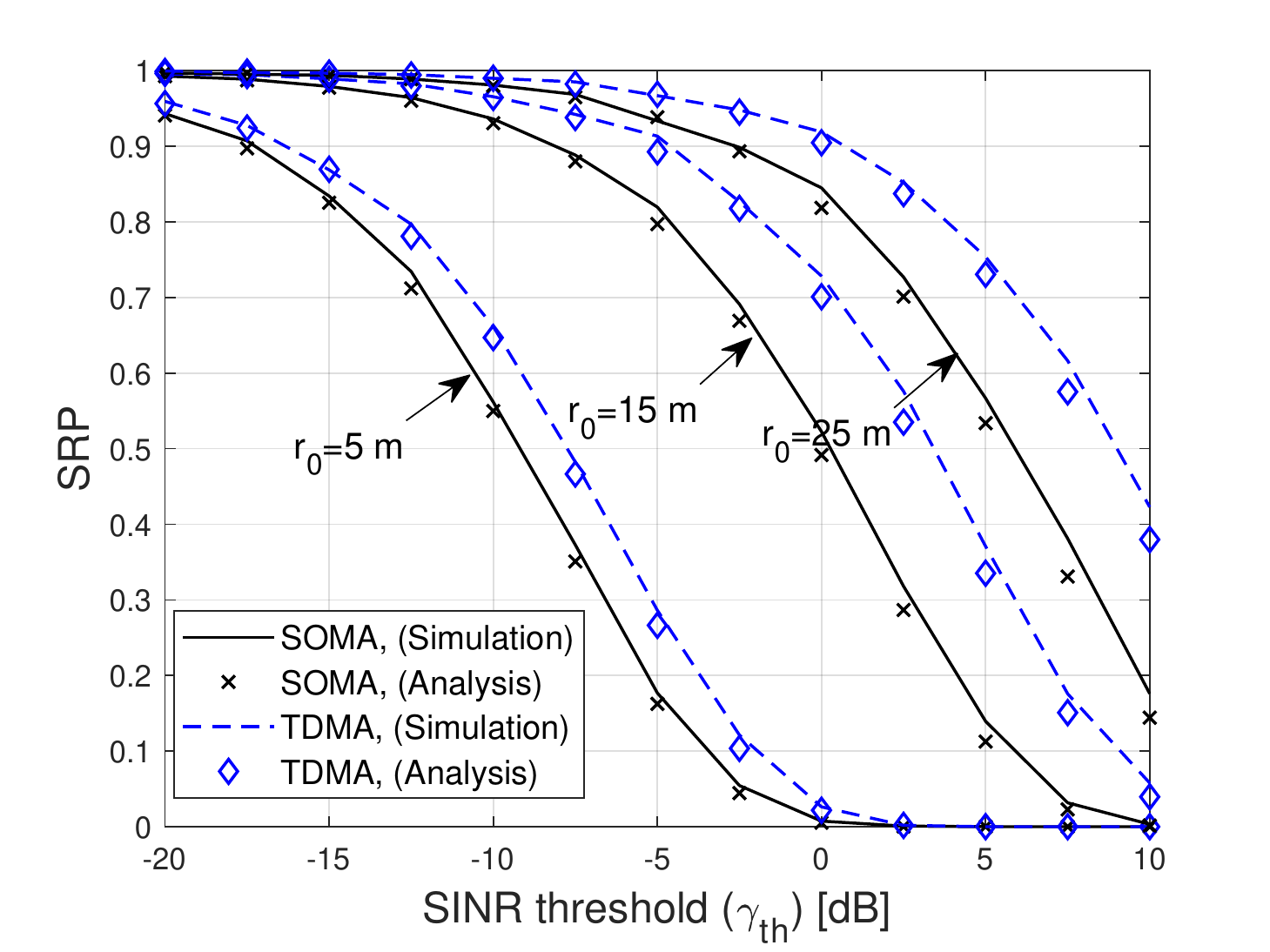}\label{fig:SRP_SINR}}
 
	\subfloat[SINR threshold vs. TC (case 1)]{\includegraphics[width=0.48\textwidth]{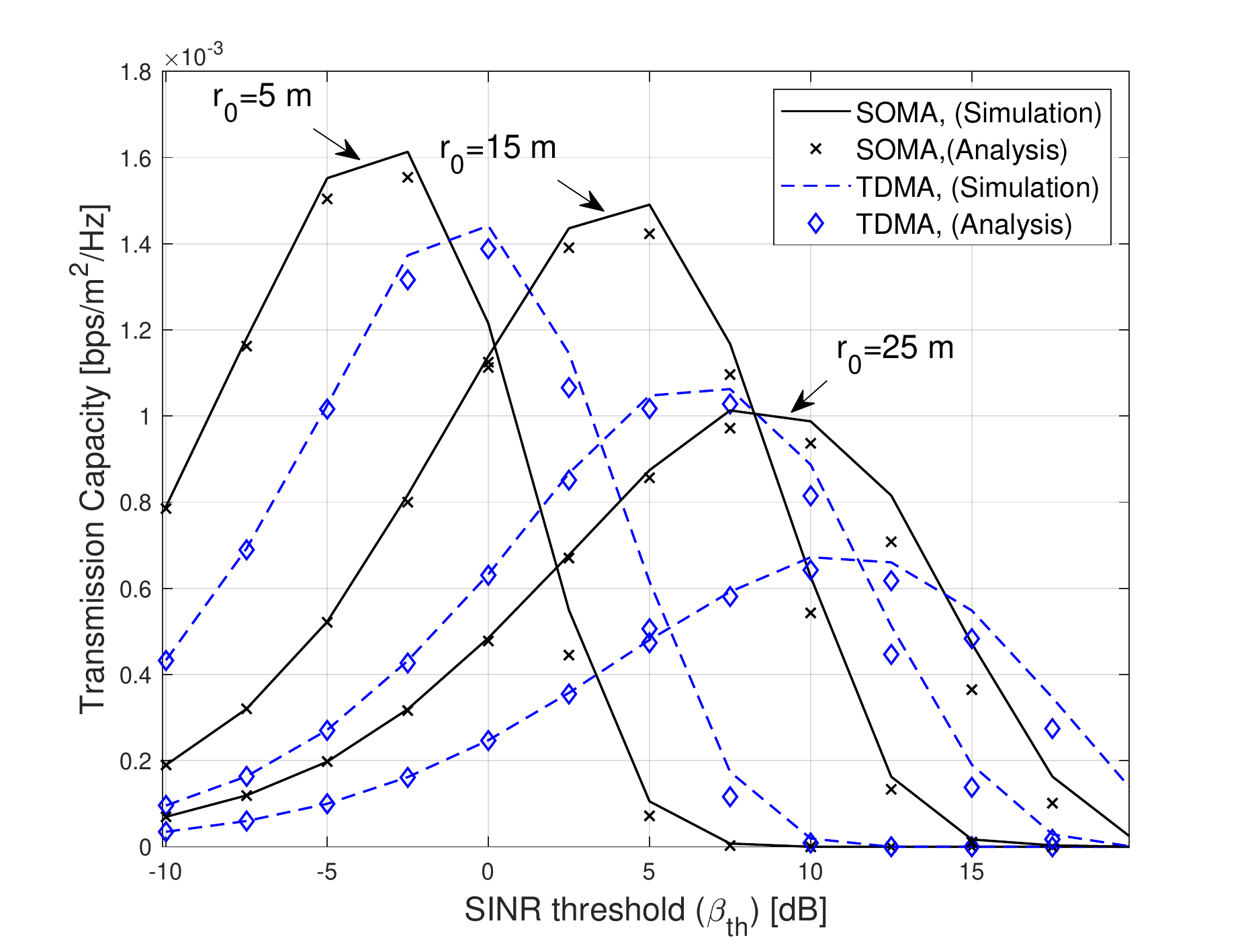}\label{fig:TC_SINR}}
	\caption{ SRP and  TC depending on  SINR threshold and  radius of  guard zone ($r_0$) where $\phi=0.5$, $\tau=0.5$, $\lambda^{\prime}_{\rm d}=0.01$, $\lambda^{\prime}_{\rm r}=0.1$. In case 1, TDMA outperforms SOMA on  SRP while  SOMA is superior to TDMA on  TC.}\label{fig:SRP_TC_SINR}
\end{figure}

\section{Simulation Results}\label{Sec:Results}
In this section, we evaluate the performance of UAV radar and communication network coexistence  based on simulation and analysis.  SRP and  TC with SOMA and TDMA are presented with the change of the different parameters. We consider 35~GHz carrier frequency for mmWave communication and Ka-band radar. The key parameters are listed in Table~III.
\subsection{SRP and TC Dependence  SINR Threshold and  Radius of  Guard Zone}

\begin{figure}[t!]
	\centering
	\subfloat[SINR threshold vs. SRP (case 2).]{\includegraphics[width=0.48\textwidth]{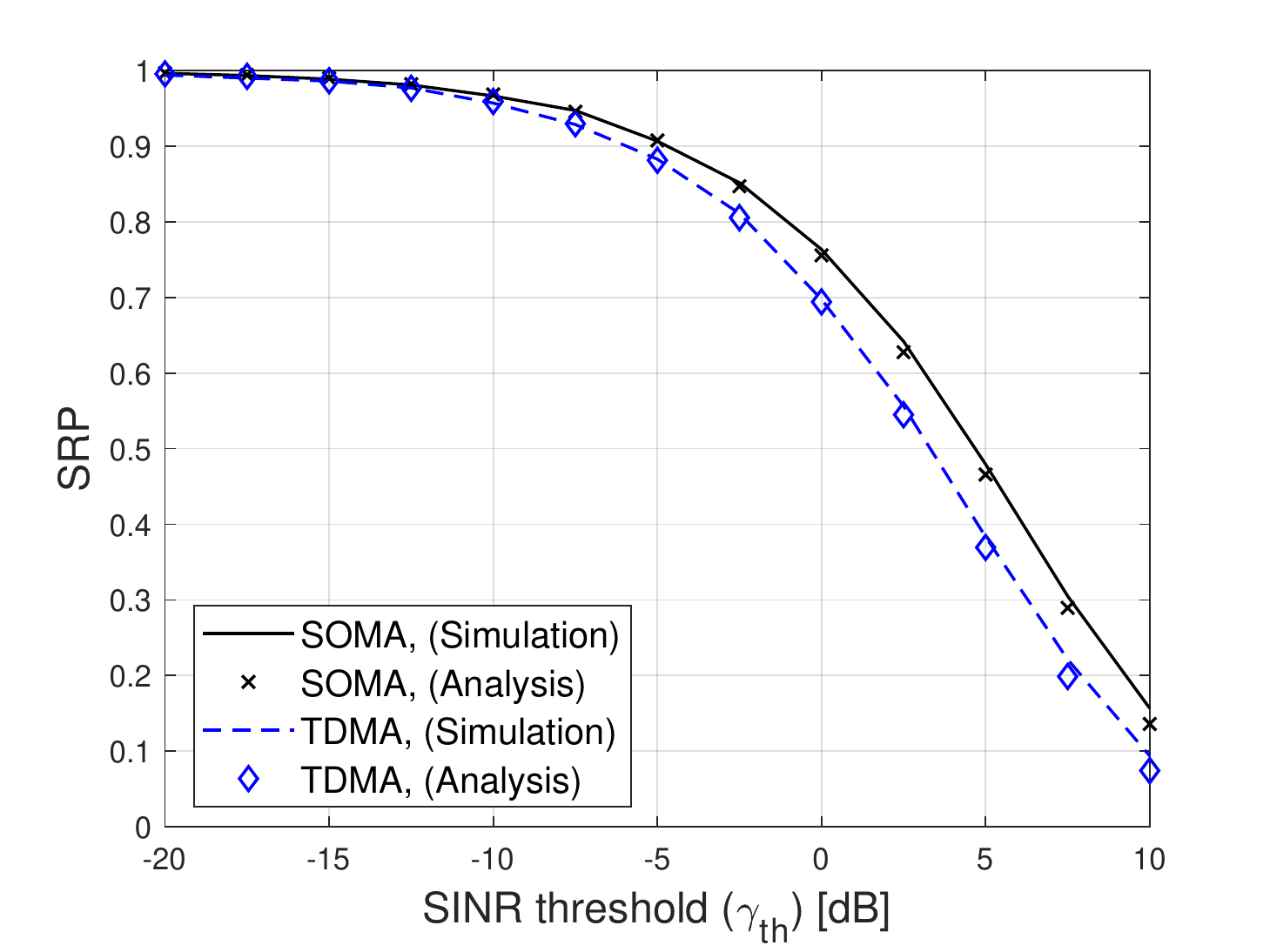}\label{fig:SRP_SINR_c2}}
 
	\subfloat[SINR threshold vs. TC (case 2).]{\includegraphics[width=0.48\textwidth]{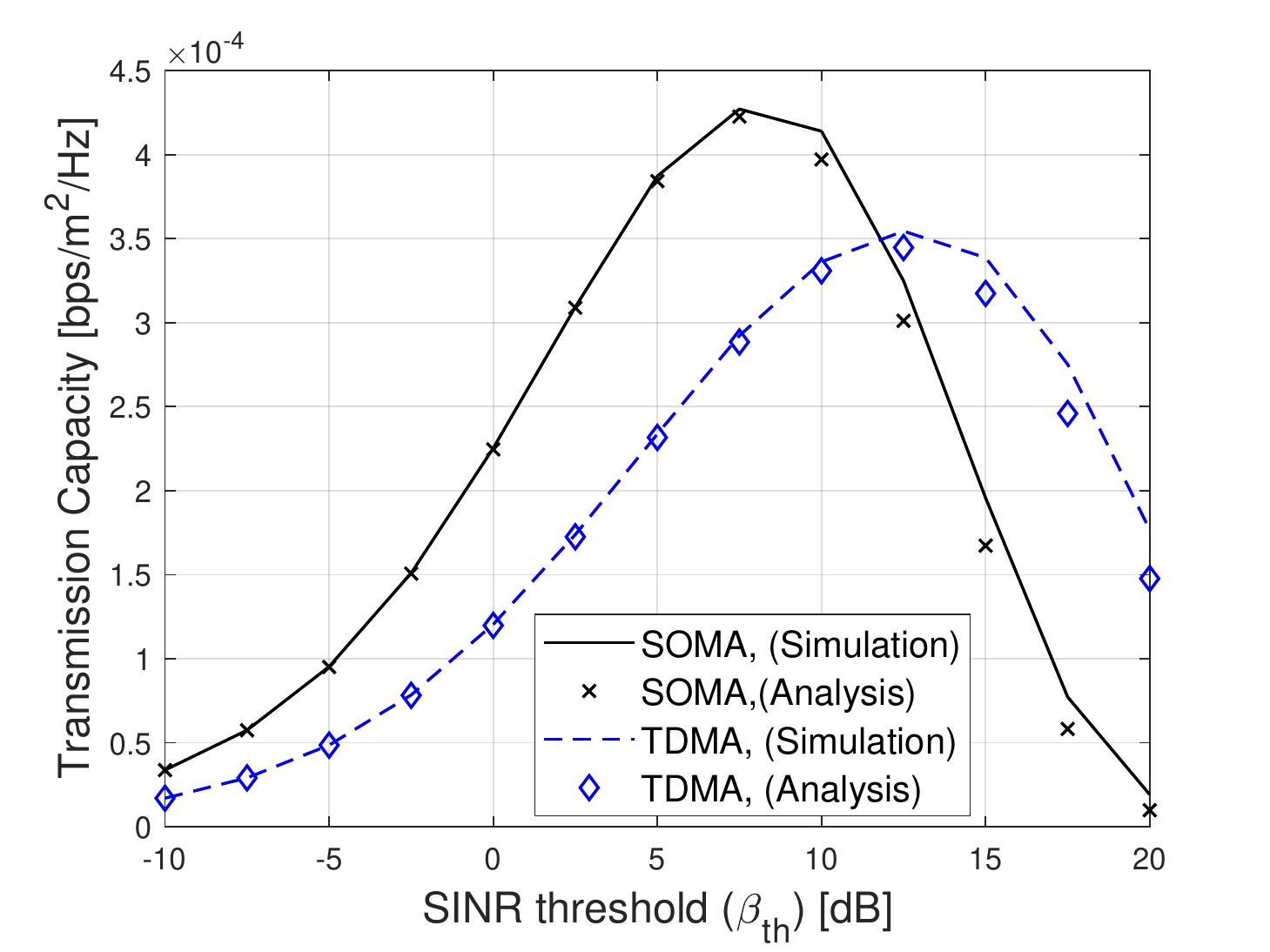}\label{fig:TC_SINR_c2}}
	\caption{The SRP and the TC depending on  SINR threshold where $\lambda^{\prime}_{\rm d}=0.00025$, $\lambda^{\prime}_{\rm r}=0.005$, $r_0=5$ m, $\phi=0.5$, $\tau=0.5$. In case 2, SOMA outperforms TDMA on both the SRP and the TC.}\label{fig:SRP_TC_SINR_c2}
\end{figure}
In this subsection, we compare SOMA and TDMA by  SRP and  TC depending on  radius of  guard zone and  SINR threshold. In Fig.~\ref{fig:SRP_TC_SINR}, we show  SRP and  TC of both SOMA and TDMA with $\lambda^{\prime}_{\rm d}$, $\lambda^{\prime}_{\rm r}$ and $\phi$, $\tau$ by case 1 in Section~\ref{sec:case 1}. As we discuss in \textit{Proposition 2} and \textit{Proposition 3}, TDMA outperforms SOMA in  SRP while SOMA is superior to TDMA in the TC. We also observe that as  radius of  guard zone $r_0$ increases,  SRP improves but  TC degrades, which is matched to the analysis in Section~\ref{sec:SRP_radius} and Section~\ref{sec:TC_radius}.

Fig.~\ref{fig:SRP_TC_SINR_c2} shows  SRP and  TC with a system configuration in case 2 in Section~\ref{sec:case 2}. It is observed that both  SRP and  TC could be better in SOMA if we consider case 2, which is mentioned in \textit{Remark 2}. Note that in a general system parameter setting, we obtain  SRP and the TC performance of case 1.

\subsection{SRP and TC Dependence  Power Splitting Factor and  Time Division Factor}
\begin{figure}[t!]
	\centering
	\subfloat[SRP vs. TC depending on $\phi$ in SOMA]{\includegraphics[width=0.48\textwidth]{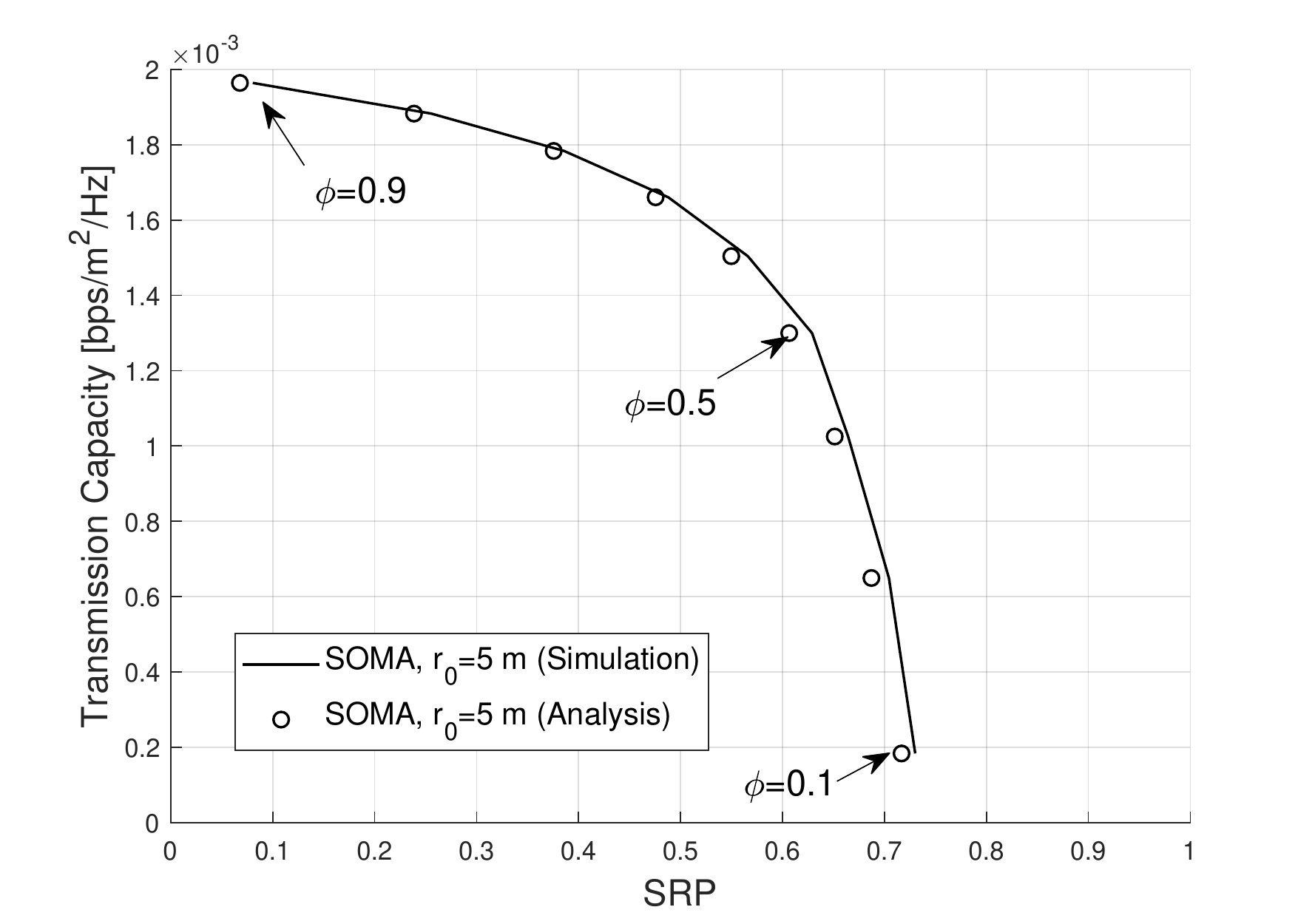}\label{fig:SRP_TC_phi}}
 
	\subfloat[SRP vs. TC depending on $\tau$ in TDMA]{\includegraphics[width=0.48\textwidth]{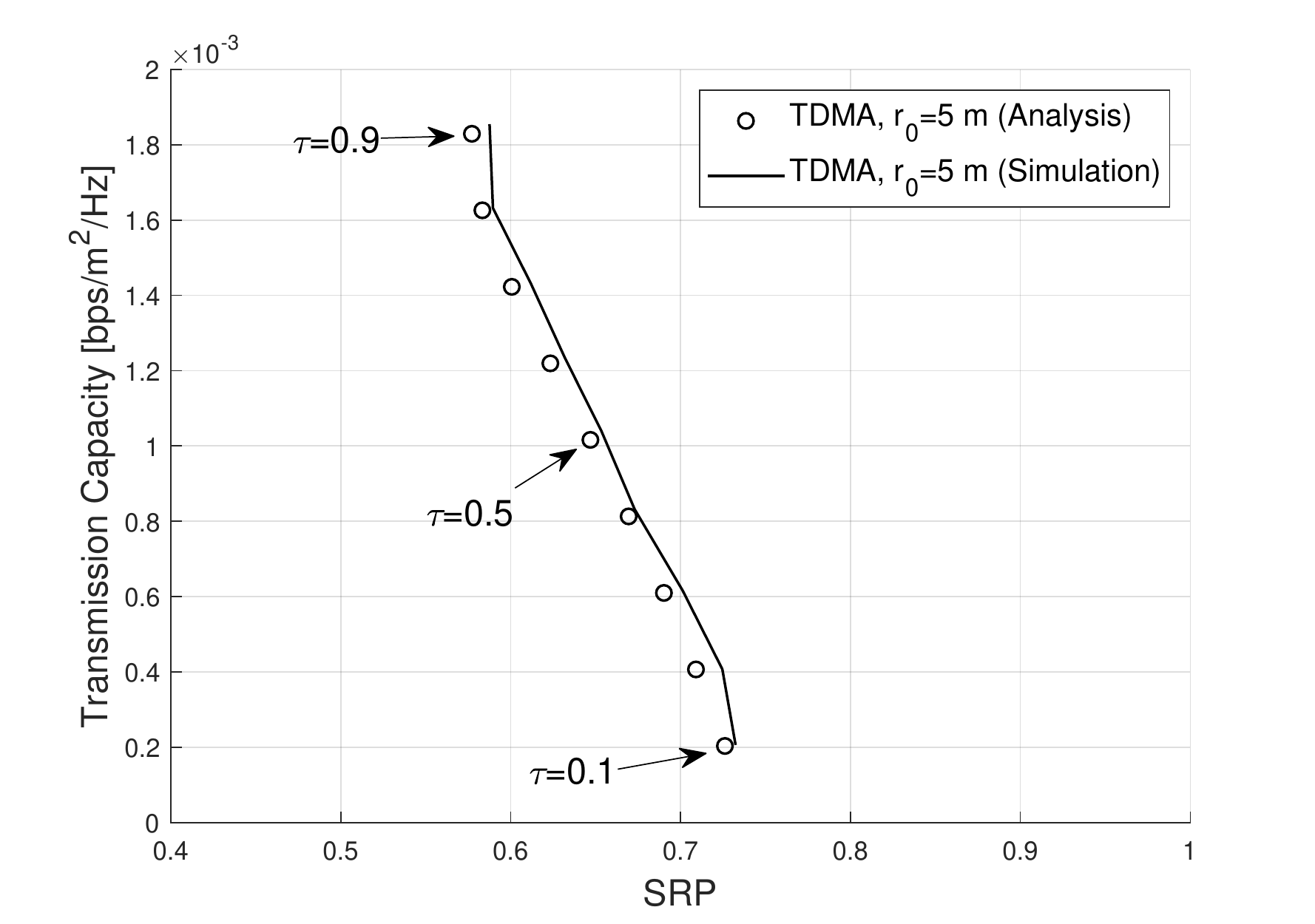}\label{fig:SRP_TC_tau}}
	\caption{Change of  SRP and  TC as $\phi$ in SOMA and $\tau$ in TDMA increases where $\lambda^{\prime}_{\rm d}=0.01$, $\lambda^{\prime}_{\rm r}=0.1$, $\beta_{\rm th}=-5$ dB, $\gamma_{\rm th}=-10$ dB.}\label{fig:SRP_TC}
\end{figure}
\begin{figure}[t!]
	\centering
	\includegraphics[width=0.48\textwidth]{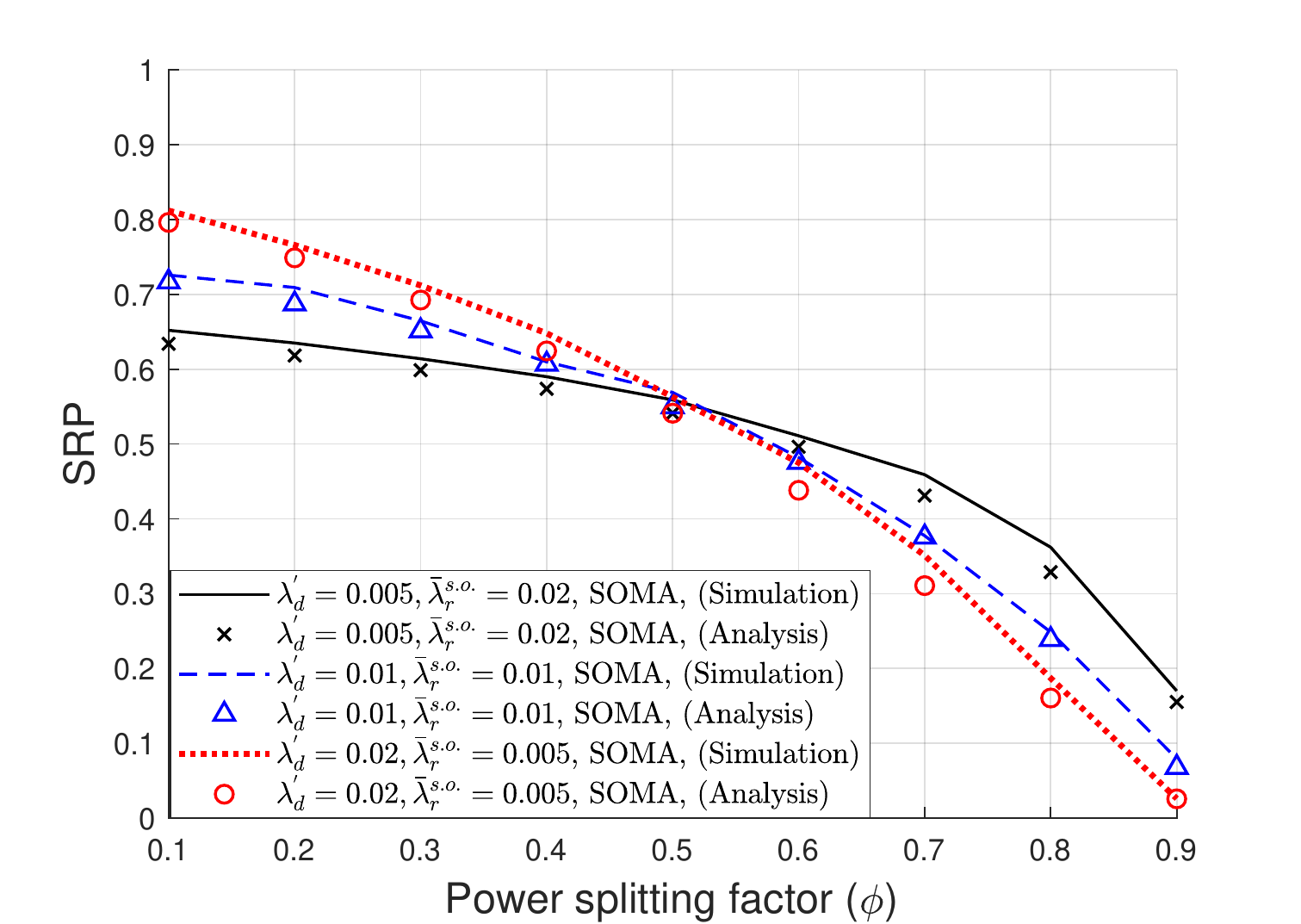}
	\caption{Change of  SRP as $\phi$ increases with different ratio of the node density of the UAV-radar and the UAV-comm where $\gamma_{\rm th}=-10$~dB, which is analyzed in \textit{Propostion 1}.}\label{fig:SRP_phi_ratio}
\end{figure}
In this subsection, we evaluate  SRP and  TC depending on $\phi$ in SOMA and $\tau$ in TDMA. Fig.~\ref{fig:SRP_TC_phi} shows that as $\phi$ increases  TC improves but  SRP decreases as we discuss in Section~\ref{sec:SRP_phi} and in Section~\ref{sec:TC_phi_tau}, which represents the impact of the different power ratio between the radar signal and the data signal on  SRP and  TC. In, Fig.~\ref{fig:SRP_TC_tau}, it is observed that  TC increases as $\tau$ becomes large. On the other hand, the SRP slowly decreases as $\tau$ increases when we compare it with $\phi$ in SOMA in Fig.~\ref{fig:SRP_TC_phi}. 

Fig.~\ref{fig:SRP_phi_ratio} show the effect of the different radio of the active UAV-radar node density $\bar{\lambda}_{\rm r}^{\rm s.o.}$ and the UAV-comm node density $\lambda_{\rm d}^{\prime}$ on the SRP in SOMA. It is observed that when $0.1<\phi<0.5$, higher UAV-comm node density achieves higher SRP  while when $0.5<\phi<1$, higher active UAV-radar node density achieves higher SRP, which can be interpreted by \textit{Proposition 1}.

\subsection{SRP and TC Dependence  Node Density of UAV-radar and UAV-comm}
\begin{figure}[t]
	\centering
	\subfloat[$\lambda_{\rm r}^{\prime}$ vs. SRP.]{\includegraphics[width=0.48\textwidth]{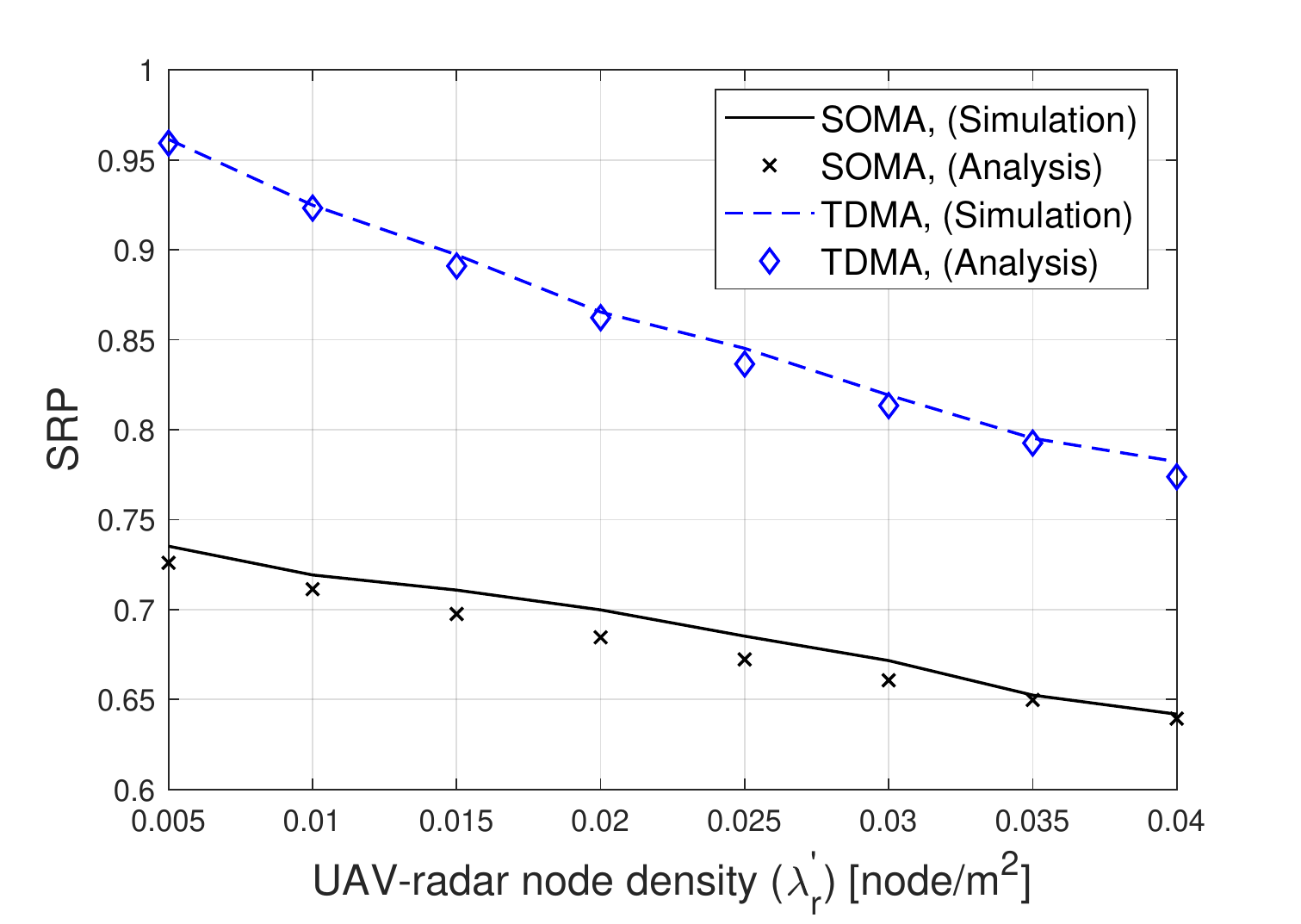}\label{fig:SRP_lammda_r}}
 
	\subfloat[$\lambda_{\rm r}^{\prime}$ vs. TC.]{\includegraphics[width=0.48\textwidth]{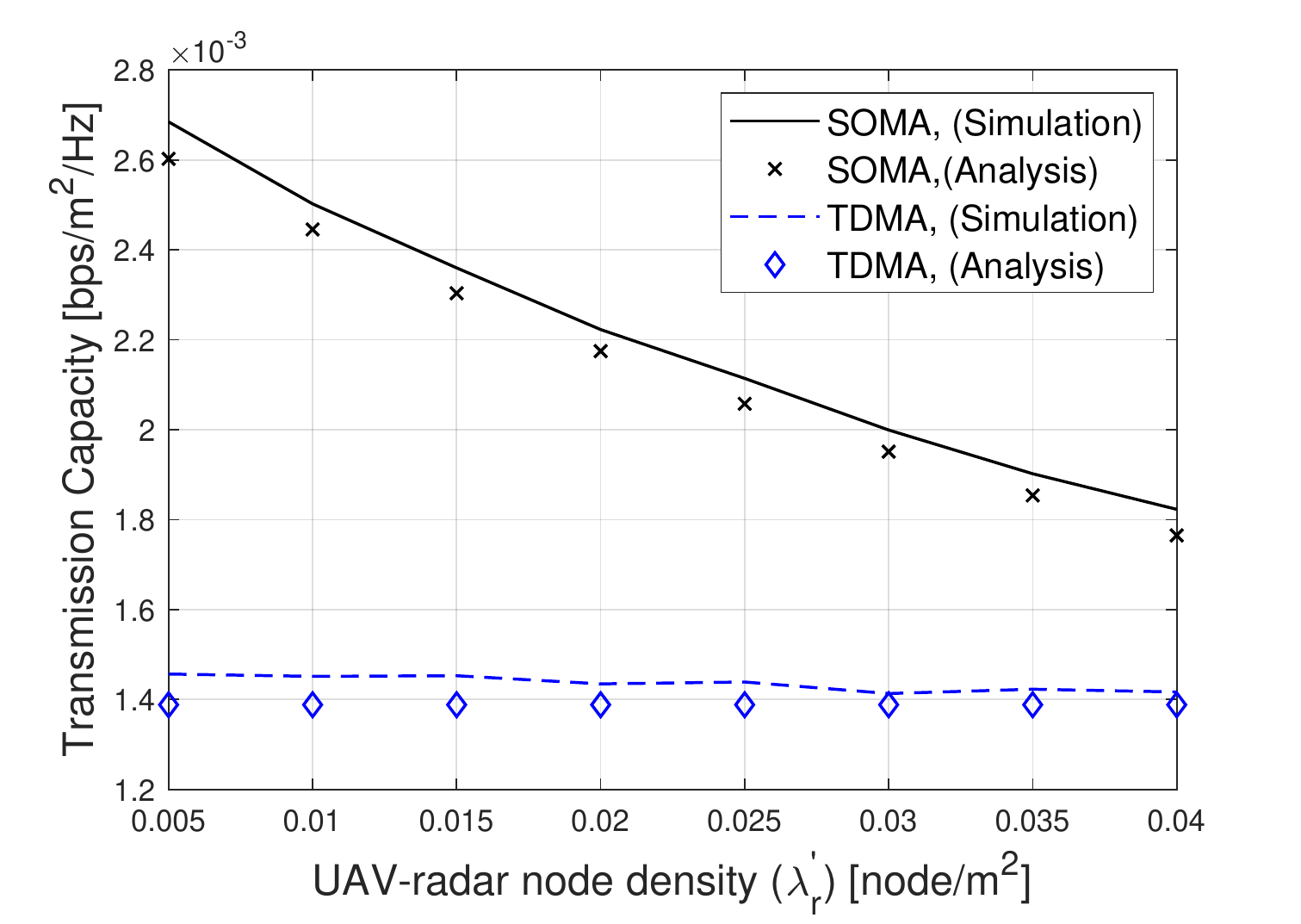}\label{fig:TC_lammda_r}}
	\caption{Change of  SRP and  TC depending on the UAV-radar node density where $\lambda^{\prime}_{\rm d}=0.01$, $r_0=5$ m, $\beta_{\rm th}=0$ dB, $\gamma_{\rm th}=-10$ dB,  $\phi=0.5$, $\tau=0.5$.}\label{fig:SRP_TC_lammda_r}
\end{figure}
\begin{figure}[t]
	\centering
	\subfloat[$\lambda_{\rm d}^{\prime}$ vs. SRP.]{\includegraphics[width=0.48\textwidth]{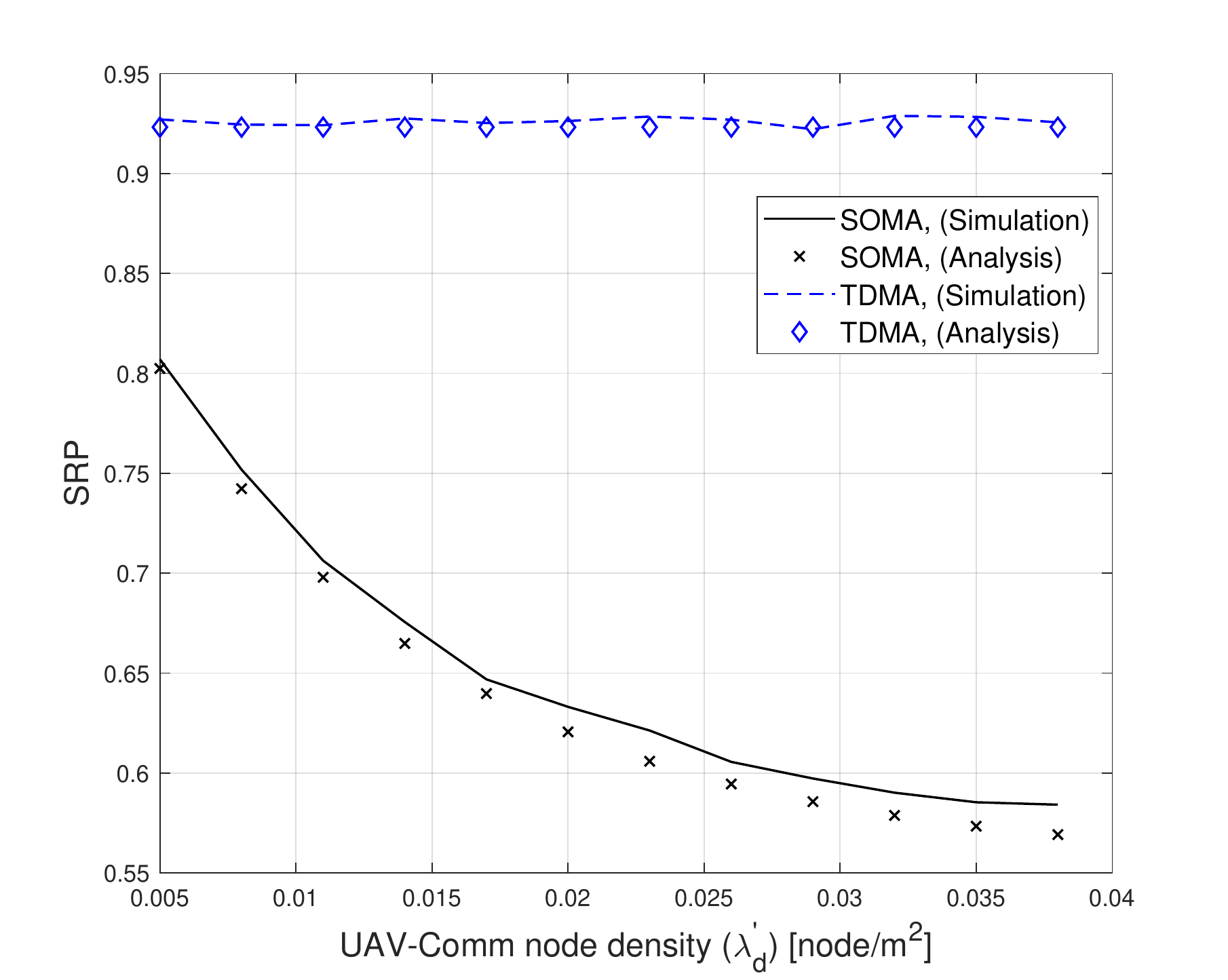}\label{fig:SRP_lammda_d}}
 
	\subfloat[$\lambda_{\rm d}^{\prime}$ vs. TC.]{\includegraphics[width=0.48\textwidth]{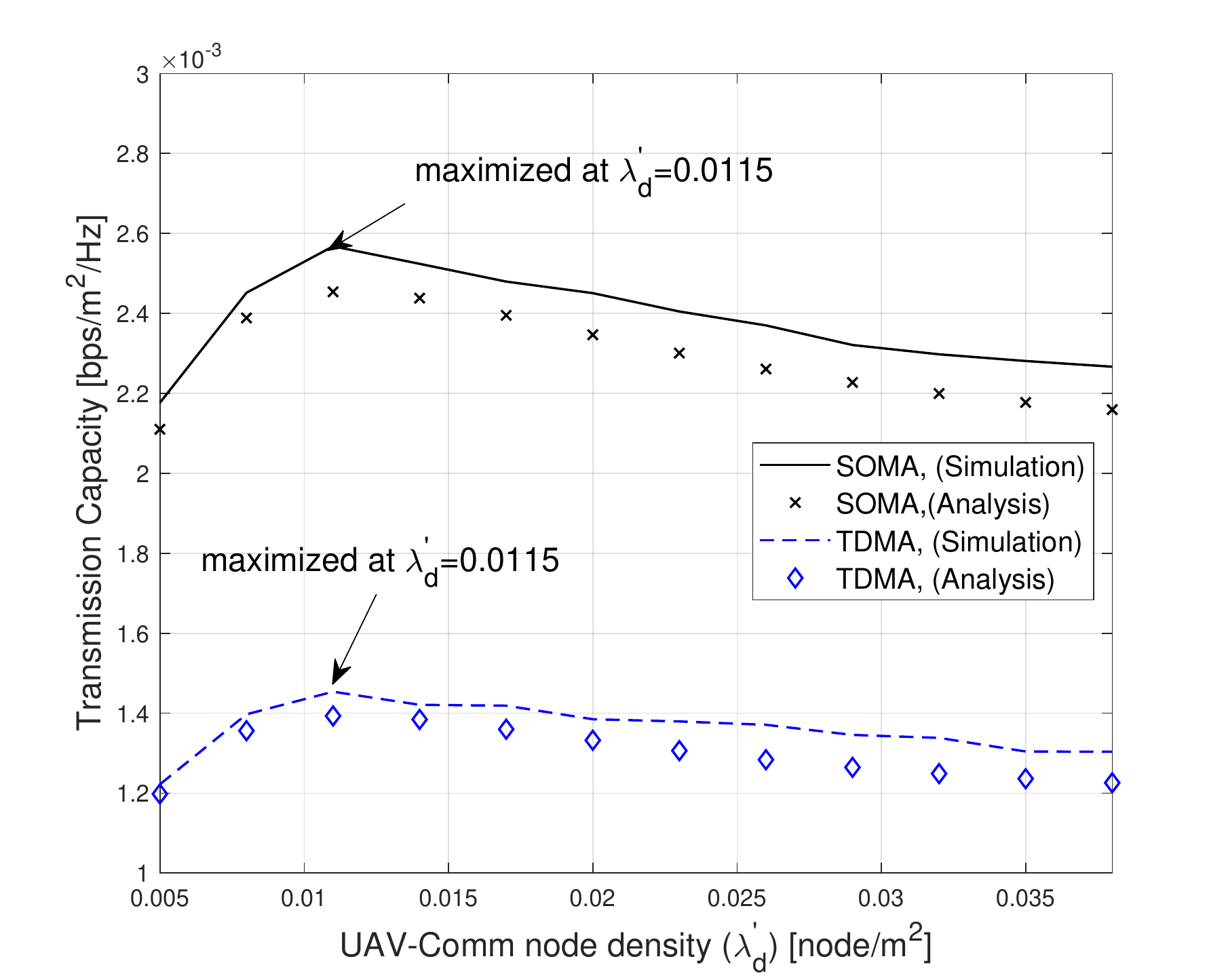}\label{fig:TC_lammda_d}}
	\caption{Change of  SRP and  TC depending on the UAV-comm node density where $\lambda^{\prime}_{\rm r}=0.01$, $r_0=5$ m, $\beta_{\rm th}=0$ dB, $\gamma_{\rm th}=-10$ dB,  $\phi=0.5$, $\tau=0.5$.}\label{fig:SRP_TC_lammda_d}
\end{figure}

In this subsection, we simulate  the dependence of SRP and  TC on the node density of the UAV-radar ($\lambda_{\rm r}^{\prime}$) and the UAV-comm ($\lambda_{\rm d}^{\prime}$). As we discuss in Section~\ref{sec:SRP_nd} and  Section~\ref{sec:TC_nd}, Fig.~\ref{fig:SRP_lammda_r} shows that  SRP is a decreasing function of $\lambda_{\rm r}^{\prime}$ for both SOMA and TDMA. In addition, in Fig.~\ref{fig:TC_lammda_r}, it is observed that  TC decreases as $\lambda_{\rm r}^{\prime}$ increases in SOMA, however, the TC is not affected by $\lambda_{\rm r}^{\prime}$ in TDMA.

In Fig.~\ref{fig:SRP_lammda_d}, we observe that  SRP is a decreasing function of $\lambda_{\rm d}^{\prime}$ in SOMA, while  SRP is not affected by $\lambda_{\rm r}^{\prime}$ in TDMA. Fig.~\ref{fig:SRP_TC_lammda_d} shows that  TC is maximized at ${\lambda}_{\rm d}^{\prime\star}=0.0115$ for both SOMA and TDMA, which can be derived from \textit{Remark~1} and \eqref{eq:nd_so}.

\section{Conclusion}\label{Sec:Conclusion}
In this paper, we investigate the coexistence of UAV radar and communication network. We deploy UAV-radars and UAV-comms by using HPPP where UAV-radars detect and track targets and UAV-comms communicate with their serving users in the same frequency band. We take into account two different multiple-access protocols, SOMA, and TDMA, to operate both radar signals and data signals simultanously. We analyze the performance of  SRP in the radar detection scenario and  TC in the data communication scenario. We show that in general, TDMA outperforms SOMA on SRP, while SOMA outperforms TDMA on TC. However, SOMA can achieve higher SRP as well as higher TC when the node density of UAV-radars is higher than that of UAV-comms (i.e., $\lambda_{\rm d}^{\prime}<\bar{\lambda}_{\rm r}$), for $\phi=\tau=0.5$, and $\mathsf{Pr}(\beta_1<\beta_{\rm th})<\frac{1}{2}$. We also find the UAV-comm node density that maximizes  TC by derving the first and the second derivative of its analytic form and analyze the behavior of SRP and TC depending on the node density, the radius of the guard zone, the power splitting factor, and  time division factor.

\bibliographystyle{IEEEtran}
\bibliography{IEEEabrv,references}

\end{document}